\documentclass[10pt, twocolumn, aps, prb]{revtex4-2}

\usepackage{graphicx} 
\usepackage{longtable, booktabs, dcolumn}
\usepackage[protrusion=true]{microtype}
\usepackage{amsmath}
\usepackage{bbold} 
\usepackage{siunitx} 
\DeclareSIUnit\angstrom{\text {Å}}
\DeclareSIUnit\rydberg{\text{Ry}}
\usepackage[version=4]{mhchem}
\usepackage{dlwf_macros}

\usepackage{hyperref} 
\hypersetup{colorlinks,
            linkcolor=blue,
            citecolor=blue,
            urlcolor=black
           }

\begin{document}
\title{Wannier Functions Dually Localized in Space and Energy}
\author{Aaron Mahler}
\affiliation{Duke University, Department of Physics, Durham, NC 27708}
\author{Jacob Z. Williams} 
\affiliation{Duke University, Department of Chemistry, Durham, NC 27708}
\author{Neil Qiang Su} 
\affiliation{Department of Chemistry, Key Laboratory of Advanced Energy Materials Chemistry (Ministry of Education) and
Renewable Energy Conversion and Storage Center (RECAST), Nankai University, Tianjin 300071, China}
\affiliation{Duke University, Department of Chemistry, Durham, NC 27708}
\author{Weitao Yang} 
\email{weitao.yang@duke.edu}
\affiliation{Duke University, Department of Chemistry, Durham, NC 27708}
\affiliation{Duke University, Department of Physics, Durham, NC 27708}

\date{\today}

\begin{abstract}
The construction of Wannier functions from Bloch orbitals offers a unitary
freedom that can be exploited to yield Wannier functions with advantageous
properties. Minimizing the spatial variance is a well-known choice; another,
previously proposed for Wannier functions constructed from the occupied Bloch
manifold, minimizes a weighted sum of spatial and energy variance. Departing
from all previous work, we extend dual localization to include both
valence and conduction bands together. Near the Fermi energy, these dually
localized Wannier functions yield frontier (bonding and antibonding) orbitals
in bulk silicon and molecular ethylene, as well as $d$-orbital character in
metallic copper. Because they are both localized and retain information about
the orbital energy spectrum, dually localized Wannier functions are well suited
to orbital-dependent methods that associate Wannier functions with specific
energy ranges. They naturally induce fractional occupations, allowing for
corrections to the DFA total energy.
\end{abstract}

\maketitle

\section{Introduction}
Choosing an orbital basis is one of the first decisions that must be made in a
single-particle electronic structure theory calculation. The eigenstates
$\ket{\psi_i}$ of the Hamiltonian are perhaps the most obvious choice, since
they separate the ground and excited states, and their associated orbital
energies $\eps_i$ are related to experimentally meaningful quantities like
ionization potentials \cite{koopmans1934, janak1978, perdew1982, cohen2008a}.
But they are not always convenient; for example, they may be spatially
delocalized. This always occurs for periodic Hamiltonians, whose eigenstates
are Bloch functions $\psi_{\bk i}(\br) = e^{i\bk\cdot\br} u_{\bk i}(\br)$, a
plane wave multiplied by a function sharing the Hamiltonian's periodicity
\cite{ashcroft1976}. Bloch functions' delocalization limits their utility in
describing spatially local properties like chemical bonds.

Wannier functions (WFs) \cite{wannier1937}, a localized, orthonormal basis set
with the same translational symmetry as the Bloch orbitals, offer a compelling
alternative. WFs were used to describe the electronic structure of crystals
\cite{kohn1959, descloizeaux1963, descloizeaux1964a} long before there was a
practical way to compute them \cite{kohn1973}. Modern uses of Wannier functions
rely on the gauge freedom, or unitary ambiguity, in their definition, which allows
considerable flexibility in their construction. The standard implementation
today is doubtless the maximally localized Wannier functions (MLWFs) of
\textcite{marzari1997}. MLWFs choose the gauge that minimizes a cost function
measuring the WFs' spatial variance; when constructed from the occupied bands of
insulators with vanishing Chern numbers, MLWFs decay exponentially in three and
fewer dimensions \cite{kohn1959, descloizeaux1964a, brouder2007, panati2013}.
MLWFs also see widespread use in electronic structure calculations; for
example, evaluating dipole moments
\cite{silvestrelli1999a, silvestrelli1999b, silvestrelli1999c}, as a basis set
for large-scale simulations 
\cite{lee2005, li2011a, shelley2011, goedecker1999}, interpolating band
structures \cite{souza2001}, and as a picture of chemical bonding
\cite{marzari1997, Resta:893386}. Localized orbitals are necessary for
corrections to delocalization error in density functional theory
\cite{mori-sanchez2008, nguyen2018}, and MLWFs are a well-studied choice
\cite{stengel2008, borghi2014, ma2016, wing2021, colonna2022}.

Despite their many successes, however, maximally localized Wannier functions cannot
serve as a complete basis for single-particle orbitals. Part of the process for
doing calculations with MLWFs is choosing the bands from which they are
constructed; in the case of semiconductors, this is often the valence manifold.
If more bands are included in the MLWFs' construction, they will become more
localized (since the Bloch functions are a complete basis). In the limit at
which all conduction bands are included, MLWFs tend to a comb of delta
functions, which are no longer useful for calculations. The recent
projectability disentanglement \cite{qiao2023b} and automated mixing
\cite{qiao2023a} methods allow MLWFs to be constructed automatically with
admixture from low-energy conduction bands, but they still cannot be used as a
complete basis. \textcite{gygi2003} proposed adding energy variance to the MLWF
cost function; this alternate gauge, described as mixed Wannier--Bloch
functions, was implemented for valence bands by \textcite{giustino2006}.

In this work, we extend the mixed spatial--energy localization gauge to the full
space of valence and conduction bands. In this context, we call the localized
orbitals dually localized Wannier functions (DLWFs). DLWFs have three unique
features. First, the addition of energy localization makes DLWFs a complete
basis for single-particle states: they can be constructed freely from any
number of valence and conduction bands. Second, mixing the valence and
conduction manifolds yields DLWFs with fractional occupations. These induce
changes $\Delta E$ to the total energy in methods to correct delocalization
error in density functional theory, such as the localized orbital scaling
correction (LOSC) \cite{li2018, su2020, mahler2022b} and the Koopmans-compliant
Wannier (KCW) functional \cite{nguyen2018, colonna2022}. A nonzero $\Delta E$ is
critical for describing the breaking of molecular chemical bonds and in
describing large, ionized molecules \cite{li2018}; therefore, it is also
necessary for the consistent description of bulk--molecule interfaces. Third,
DLWFs whose energy is localized near the Fermi level appear as localized
frontier orbitals, capturing both bonding and antibonding character in
silicon and ethylene as well as orbitals resembling the $d$ manifold in copper.
They extend the frontier molecular orbitalet paradigm \cite{yu2022a} to
materials, a necessary step toward simulating chemical reactivity at interfaces.
DLWFs thus promise broad application to simulating electronic structure and
chemical reactions of both materials and interfaces.

\section{Theory} \label{sec:methods}
We will assume a Monkhorst--Pack mesh \cite{monkhorst1976}, in
which a uniform grid of $N_k$ $\bk$-points (including the origin $\Gamma$) in
the first Brillouin zone define the Bloch orbitals, throughout. Under this
assumption, Wannier functions take the form
\begin{equation} \label{eqn:canonicalWF}
\ket{w_{\bR i}} = \frac{1}{N_k} \sum_{\bk} e^{-i\bk\cdot\bR} \ket{\psi_{\bk i}},
\end{equation}
where $\bR$ indexes unit cells in the unfolded supercell on which the Wannier
functions are periodic (following Born--von K\`arm\`an boundary conditions
\cite{ashcroft1976}). This normalization convention implies that 
$\inn{\psi_{\bk i}}{\psi_{\bq j}} = N_k \delta_{ij} \delta_{\bk\bq}$ and that
$\ket{\psi_{\bk i}} = \sum_{\bR} e^{i\bk\cdot\bR} \ket{w_{\bR i}}$. For
simplicity of notation, we omit the spin index, but our arguments apply
independently to each spin channel of a collinear spin-polarized system.

\subsection{Gauge freedom}
At each $\bk$, Bloch orbitals are defined only up to a global phase
$e^{i\theta_{\bk n}}$. This yields a gauge freedom when constructing a Wannier
functions from a single band, since
\begin{equation}
    \ket{w_{\bR i}} =
    \frac{1}{N_k} \sum_{\bk}
        e^{-i\bk\cdot\bR} e^{i\theta_{\bk i}} \ket{\psi_{\bk i}}.
\end{equation}
To construct localized WFs, $\theta_i(\bk)$ should be chosen such that
$e^{i\theta_i(\bk)} \psi_i(\bk)$ is as smooth as possible; if it is analytic,
the resulting Wannier function is exponentially localized
\cite{descloizeaux1964a}. When Wannier functions are constructed from multiple
bands, the gauge freedom expands to include any unitary combination $U^{\bk}$
of the bands at the same $\bk$-point \cite{descloizeaux1963}. (These are
formally called \emph{generalized} Wannier functions, but we will refer to them
as Wannier functions throughout the text.) Thus
\begin{equation}
\begin{split}
    \ket{w_{\bR i}} &= 
    \frac{1}{N_k} \sum_{\bk} e^{-i\bk\cdot\bR} 
        \sum_j U^{\bk}_{ji} \ket{\psi_{\bk j}} \\ &=
    \frac{1}{N_k} \sum_{\bk} e^{-i\bk\cdot\bR} \ket{\phi_{\bk i}}.
\end{split}
\end{equation}
We call the intermediate
$\ket{\phi_{\bk i}} = \sum_j U^{\bk}_{ji} \ket{\psi_{\bk j}}$ transformed
Bloch orbitals.

\subsection{Cost functions} \label{sec:costFuncs}
It is the unitary gauges $U^{\bk}$ that give Wannier functions their practical
flexibility. Maximally localized Wannier functions \cite{marzari1997} choose
$U^{\bk}$ that minimize the cost function
\begin{equation} \label{eqn:spatialCost}
\Omega = \sum_i \mel{w_{\bZ i}}{\Delta r^2}{w_{\bZ i}}; \\
\end{equation}
that is, the sum of each MLWF's spatial variance
\begin{equation}
    \evt{\Delta r^2}{w_{\bZ i}} = 
    \evt{r^2}{w_{\bZ i}} - \abs{\evt{\br}{w_{\bZ i}}}^2,
\end{equation}
integrated over the supercell. Eq.~(\ref{eqn:spatialCost}) is the natural
extension to periodic boundary conditions of Foster--Boys localization for
finite systems \cite{foster1960}. Without loss of generality, only the home
unit cell $\bR = \bZ$ is included in the cost function because Wannier functions
at $\bR$ are related by translation to those at $\bZ$,
$w_{\bR i}(\br) = w_{\bZ i}(\br - \bR)$. Summing over $\bR$ would thus change
$\Omega$ only by a multiplicative constant. The MLWF formulation was originally
applied to a \emph{composite} set of Bloch bands, separated by an energy gap
from all other states at every point in the Brillouin zone. The gradient of
$\Omega$ was derived analytically for a composite set by \citet{marzari1997},
and descent methods are applied to obtain the $U^{\bk}$ which minimize
$\Omega$. 

\textcite{gygi2003} proposed adding an energy variance term to $\Omega$,
yielding the dual-localization cost function
\begin{multline} \label{eqn:L2Cost}
    F \coloneqq
    (1 - \gamma) \Omega +C \gamma \Xi \\ =
    (1 - \gamma) \sum_i \mel{w_{\bZ }}{\Delta r^2}{w_{\bZ i}} \\ +
        C \gamma \sum_i \mel{w_{\bZ i}}{\Delta h^2}{w_{\bZ i}}.
\end{multline}
Here, $\Ham$ is the single-particle Hamiltonian, $\Omega$ the spatial spread, and
$\Xi$ the energy spread; 
\begin{equation}
    \evt{\Delta \Ham^2}{w_{\bZ i}} =
    \evt{\Ham^2}{w_{\bZ i}} - \abs{\evt{\Ham}{w_{\bZ i}}}^2
\end{equation}
is the energy variance of $\ket{w_{\bZ i}}$; and the unitful constant
$C = \qty{1}{\angstrom^2/\electronvolt^2}$  restores dimensional consistency of $F$
\footnote{The numerical value of $C$ is arbitrary; the units given here are those
used in the implementation we describe below.}.
The mixing term
$\gamma \in [0, 1]$ can be tuned to prioritize spatial or energy localization
(see Sec.~\ref{sec:dlwf.gamma}).

Our work extends \cite{gygi2003} and \cite{giustino2006}, which uses $F$ to
construct electrons and phonon WFs in bulk systems, in two ways. First,
\cite{gygi2003, giustino2006} apply only to calculations in which the Brillouin
zone is sampled only at the origin $\Gamma$ of reciprocal space. In this case,
the transformed Bloch orbitals are equivalent to the Wannier functions; varying
the mixing term $\gamma$ interpolates between MLWFs (at $\gamma = 0$) and Bloch
orbitals ($\gamma = 1$), so the resulting $\ket{w_{\bR i}}$ are called
Wannier--Bloch functions. When $N_k > 1$, a nontrivial discrete Fourier
transform is required to map $\ket{\phi_{\bk i}}$ to $\ket{w_{\bR i}}$, and the
Wannier functions do not reduce to the Bloch orbitals at $\gamma = 1$; in this
case, we call the WFs minimizing $F$ dually localized Wannier functions. Second,
Wannier--Bloch functions are constructed from only the occupied (valence) Bloch
manifold. We show that $F$ allows Wannier functions to be built from both
valence and conduction bands at once, treating both on the same footing.

\subsubsection{Minimizing the cost function}
\textcite{marzari1997} minimize $\Omega$ for a composite set of energy bands by
determining its gradient with respect to $U^{\bk}$. This is accomplished by
splitting the cost function into two positive definite quantities,
$\Omega = \Omega_{\text{I}} + \wt{\Omega}$. $\Omega_{\text{I}}$ is invariant to
the choice of gauge, so minimizing $\Omega$ is equivalent to minimizing only the
gauge-dependent term $\wt{\Omega}$. For a set of $N_w$ Wannier functions,
\begin{multline}
    \Omega_\text{I} =
    \sum_i^{N_w} 
        \Bigg( 
            \mel{w_{\bZ i}}{r^2}{w_{\bZ i}} \\ - 
            \sum_{j}^{N_w} \sum_{\bR}^{N_k}
            \abs{\mel{w_{\bR j}}{\br}{w_{\bZ i}}}^2 
        \Bigg),
\end{multline}
and the gauge-dependent spatial cost is
\begin{multline}
    \wt{\Omega} =
    \sum_{j \neq i}^{N_w} \sum_{\bR}^{N_k}
        \abs{\mel{w_{\bR j}}{\br}{w_{\bZ i}}}^2 \\ +
    \sum_i^{N_w} \sum_{\bR \neq \bZ}^{N_k}
        \abs{\mel{w_{\bR i}}{\br}{w_{\bZ i}}}^2.
\end{multline}

By an argument exactly analogous to that in \cite{marzari1997}, the energy cost
$\Xi$ splits into a gauge-invariant term $\Xi_{\text{I}}$ and the gauge-dependent
$\wt{\Xi}$,
\begin{multline}
    \Xi = \Xi_{\text{I}} + \wt{\Xi} \\ =
    \sum_i
        \left(
            \evt{\Ham^2}{w_{\bZ i}} - \sum_{\bR j}
                \abs{\mel{w_{\bR j}}{\Ham}{w_{\bZ i}}}^2
        \right) \\ +
    \sum_i \sum_{\bR j \neq \bZ i} \abs{\mel{w_{\bR j}}{\Ham}{w_{\bZ i}}}^2.
\end{multline}
Thus, for a composite set of bands, we find the unitary operators $U^{\bk}$
defining the DLWF gauge by minimizing
\begin{equation}
    \wt{F} = (1 - \gamma) \wt{\Omega} + C \gamma \wt{\Xi}.
\end{equation}
Because differentiation is linear, the gradients of $\wt{\Omega}$ and
$\wt{\Xi}$ with respect to $U^{\bk}$ can be found in parallel, and the minimum
found with an iterative procedure. Virtual Bloch bands almost never form a
composite set, however, so we require an additional step to construct DLWFs.

\subsection{Disentanglement}
The valence bands of metals, and the conduction bands of most systems, cannot be
separated from the rest of the bands by an energy gap everywhere in the
Brillouin zone; such bands are said to be \emph{entangled}. Souza et al.
developed a \emph{disentanglement} method, extracting a subset of interest from
a set of entangled bands \cite{souza2001}. From $N_b$ Bloch bands, which we now
index with $n$, disentanglement obtains $N_w \leq N_b$ composite bands that are
as smooth as possible in $\bk$; this turns out to be accomplished by minimizing
$\Omega_{\text{I}}$ iteratively. We write $\ket{\wt{\psi}_{\bk i}}$, with $1 \leq i \leq N_w$, for the disentangled bands; they are obtained as
\begin{equation}
    \ket{\wt{\psi}_{\bk i}} = \sum_n V^{\bk}_{in} \ket{\psi_{\bk n}}.
\end{equation}
Here, $V^{\bk}$ is a $N_w \times N_b$ semiunitary transformation that satisfies
$V^{\bk}(V^{\bk})^\dagger = \mathbb{1}$, the $N_w$-dimensional identity
operator. Applying $U^{\bk}$ to $\ket{\wt{\psi}_{\bk i}}$ generates the
transformed Bloch orbitals $\ket{\phi_{\bk i}}$, which are Fourier transformed
to generate $\ket{w_{\bR i}}$. The combined disentanglement, localization,
and Wannierization process is 
\begin{equation}
\begin{split}
    w_{\bR i}(\br) &=
    \frac{1}{N_k} \sum_{\bk}^{N_k}
        e^{-i\bk\cdot\bR} \phi_{\bk i}(\br) \\ &=
    \frac{1}{N_k} \sum_{\bk}^{N_k}
        e^{-i\bk\cdot\bR} \sum_j^{N_w} 
            U^{\bk}_{ij} \wt{\psi}_{\bk j}(\br) \\ &=
    \frac{1}{N_k} \sum_{\bk}^{N_k}
        e^{-i\bk\cdot\bR} \sum_j^{N_w}
            U^{\bk}_{ij} \sum_n^{N_b} 
                V^{\bk}_{jn} \psi_{\bk n}(\br).
\end{split}
\end{equation}

In principle, we could modify the disentanglement procedure in the same way
that $F$ modifies $\Omega$, finding the $N_w$-dimensional subspace that
minimizes
\begin{equation}
    F_{\text{I}} = (1 - \gamma) \Omega_{\text{I}} + \gamma \Xi_{\text{I}}.
\end{equation}
In practice, however, we find this unnecessary; we use the space-only
disentanglement of \cite{souza2001} as the first step in computing the DLWFs. In
particular, this procedure projects the Hamiltonian onto the subspace spanned by
$\{\ket{\wt{\psi}_{\bk i}}\}$ after each iteration, yielding a disentangled
Hamiltonian $\wt{h} = \sum_{\bk} V^{\bk} h (V^{\bk})^\dagger$ that satisfies
\begin{equation}
    \mel{\wt{\psi}_{\bk i}}{\wt{\Ham}}{\wt{\psi}_{\bq j}} =
    N_k \delta_{\bk\bq} \delta_{ij} \wt{\eps}_{\bk i},
\end{equation}
where $\wt{\eps}_{\bk i}$ is a disentangled energy eigenvalue. (If we modify
$\Xi$ to use the disentangled Hamiltonian, then $\Xi_{\text{I}} = 0$ trivially.
See the Supplemental Material \footnote{See the Supplemental Material at [URL
will be inserted by publisher] for further details on energy disentanglement;
detailed tables and band structures from the DLWFs studied in this work; a
discussion of the units of the space-energy mixing parameter $\gamma$; and the
modifications we made to \texttt{wannier90}. Includes reference
\cite{tiesinga2020}.} for the derivation.)

\begin{figure}[ht]
\centering
\includegraphics[width=\linewidth]{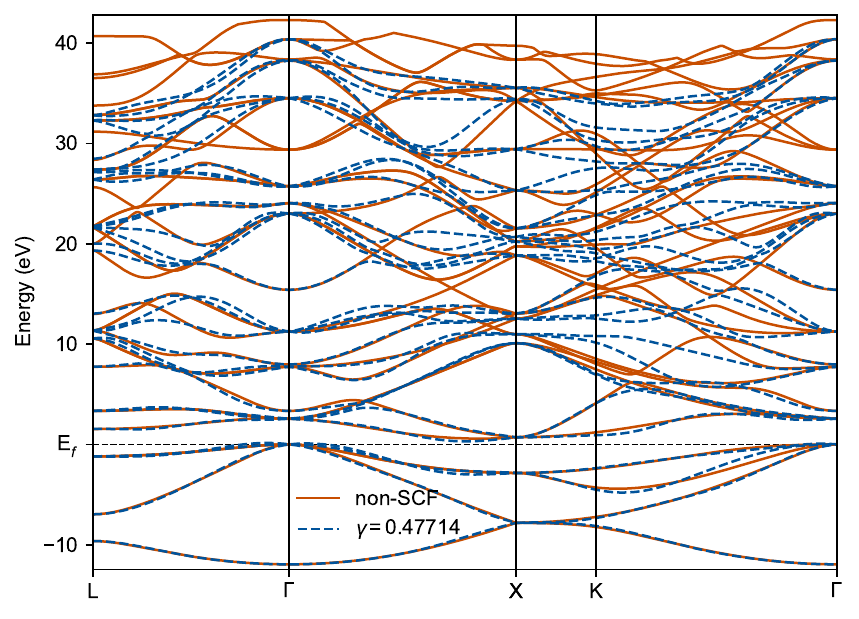}
\caption{PBE band structure of silicon (valence band maximum set to
         zero), disentangling 34 bands to 30 Wannier functions, with 
         disentangled states below the Fermi energy fixed to reproduce
         the Bloch bands exactly. Solid orange bands are
         non-self-consistently computed in \texttt{Quantum ESPRESSO};
         dashed blue bands are interpolated from the DLWFs. Mixing
         parameter $\gamma = 0.47714$.
        }
\label{fig:si.dis.b34w30}
\end{figure}

The difference between the original and disentangled Hamiltonian depends on the
subspace projectors $V^{\bk}$. If every disentangled eigenvalue
$\wt{\eps}_{\bk i}$ corresponds exactly to a Bloch eigenvalue $\eps_{\bk n}$ at
the same $\bk$-point, then $\Ham$ and $\wt{\Ham}$ commute inside the
disentangled subspace and using $\wt{\Ham}$ is equivalent to using $\Ham$ in
$\Xi$. We do not achieve this in practice, but the existing disentanglement
procedure creates a band structure nearly equivalent to the full band structure
except for high-energy conduction bands. We can thus reproduce the band
structure around the Fermi energy accurately by including sufficiently many
unoccupied orbitals before disentanglement; the band structure interpolated from
the resulting Wannier functions differs appreciably from the original only well
above the Fermi energy. For example, a band structure for silicon disentangling
34 bands to 30 yields DLWFs that interpolate a qualitatively accurate band 
structure up to \SI{20}{\electronvolt} above the Fermi energy
(Fig.~\ref{fig:si.dis.b34w30}).

Replacing $\Ham$ by the disentangled Hamiltonian $\wt{\Ham}$ also means that
$\sum_i \evt{\wt{\Ham}^2}{w_{\bZ i}}$ is gauge-invariant, since
\begin{equation}
    \sum_i^{N_w} \evt{\wt{\Ham}^2}{w_{\bZ i}} =
    \frac{1}{N_k} \tr{\left[(\wt{\Ham})^2\right]}.
\end{equation}
We thus decompose
\begin{equation}
\begin{split}
    \Xi &=
    \sum_i \left[
        \evt{\wt{h}^2}{w_{\bZ i}} - \abs{\evt{\wt{\Ham}}{w_{\bZ i}}}^2
    \right] \\ &\coloneqq
    \Xi_{\SA} - \Xi_{\AS},
\end{split}
\end{equation}
where the squared average energy term $\Xi_{\SA}$ is invariant to the choice of
Wannier basis. As is true for $\Omega$ in $\Gamma$-sampled systems, minimizing
$\Xi$ (or $\wt{\Xi}$) is equivalent to maximizing the average energy squared
term $\Xi_{\AS}$.

\subsection{Energy cost gradient} \label{sec:egrad}
We now derive the gradient of $\Xi_{\AS}$ with respect to the localization
unitary $U^{\bk}$. To first order, $U^{\bk}$ can be written as a small
anti-Hermitian perturbation to the identity;
$U^{\bk} \approx \mathbb{1} + W^{\bk}$, with $(W^{\bk})^\dagger=-W^{\bk}$. Then
$\partial \Xi_{\AS} / \partial U^{\bk}$ is obtained via the matrix calculus
identity
\begin{equation} \label{eqn:A}
    \frac{d \Re{\tr{[MW]}}}{dW} =
    \frac12 \left(M - M^\dagger\right) \coloneqq
    \mathcal{A}[M].
\end{equation}
To see why this is useful, observe that we may write
\begin{equation}
\begin{split}
    \Xi_{\text{AS}} &= 
    \sum_i \abs*{\evt{\wt{h}}{w_{\bZ i}}}^2 =
    \sum_i \abs*{\frac{1}{N_k} \textstyle\sum_{\bk j} 
        \abs{U^{\bk}_{ij}}^2 \wt{\eps}_{\bk j}}^2 \\ &=
    \sum_i \abs*{\frac{1}{N_k} \textstyle\sum_{\bk} B^{\bk}_{ii}}^2,
\end{split} 
\end{equation}
where
$B^{\bk}_{ij} = 
\begin{pmatrix} U^{\bk} \wt{\Ham} (U^{\bk})^\dagger \end{pmatrix}_{ij}$ is the Hamiltonian in the transformed Bloch basis. The change in $\Xi_{\text{AS}}$ to
first order in $W^{\bk}$ is then
\begin{equation}
\begin{split}
    d \Xi_{\AS} &= 
    \frac{4}{N_k^2} \sum_i 
        \left(
            \sum_{\bk} \Re{\begin{pmatrix} B^{\bk} W^{\bk} \end{pmatrix}_{ii}}
        \right)
        \left(
            \sum_{\bk'} B^{\bk'}_{ii}
        \right) \\ &=
    \frac{4}{N_k^2} \sum_{\bk\bk'} 
        \Re{\Tr{\left[C^{(\bk,\bk')} W^{\bk} \right]}},
\end{split} 
\end{equation}
where
$\begin{pmatrix} C^{(\bk,\bk')} \end{pmatrix}_{ij} = 
B^{\bk}_{ji} B^{\bk'}_{jj}$. We can thus write the gradient in terms of the
superoperator $\mathcal{A}$, defined in Eq.~(\ref{eqn:A}), as
\begin{equation}
\begin{split}
    \frac{d\Xi_{\AS}}{dW^{\bk}} &=
    \frac{4}{N_k} \sum_{\bk'} \mathcal{A}\left[C^{(\bk, \bk')}\right] \\ &=
    \frac{2}{N_k} \sum_{\bk'} \left[ 
        C^{(\bk, \bk')} - (C^{(\bk, \bk')})^\dagger
    \right].
\end{split} 
\end{equation}
In this form, it appears the gradient $\bk$-point depends on every other
$\bk$-point in the Brillouin zone. However, we can sum $C^{(\bk,\bk')}$ over
$\bk'$ prior to evaluating the gradient. Thus
\begin{equation} \label{eqn:dxi}
    \begin{bmatrix} \frac{d\Xi_{\AS}}{dW^{\bk}} \end{bmatrix}_{ij} =
    2 B^{\bk}_{ij}
        \left( \evt{\wt{\Ham}}{w_{\bZ j}} - \evt{\wt{\Ham}}{w_{\bZ i}} \right).
\end{equation}
Thus, $\Xi$ essentially penalizes mixing of each pair of Bloch orbitals in
proportion to the difference in average energy between the Wannier functions
they are used to construct. Eq.~\eqref{eqn:dxi} also makes clear that the
computational cost of computing $d\Xi_{\text{AS}}/dW^{\bk}$ is independent of
$N_k$, so the overall cost of energy localization is linear in $N_k$.

\subsection{Space-energy mixing parameter} \label{sec:dlwf.gamma}
DLWFs are defined by the choice of the mixing parameter $\gamma$ in
Eq.~\eqref{eqn:L2Cost}. Setting $\gamma = 0$ recovers the MLWF cost function
$\Omega$. The Bloch orbitals are recovered at $\gamma = 1$ if the Brillouin zone
is sampled only at $\Gamma$; when $N_k > 1$, setting $\gamma = 1$ yields 
transformed Bloch orbitals $\ket{\phi_{\bk n}}$ equivalent to the original Bloch
orbitals, ordered by their energy at each $\bk$. If there are band crossings in
the Brillouin zone, the energy-ordered Bloch orbitals will not be smooth in
$\bk$, giving poor spatial localization. Choosing $\gamma$ strictly between 0
and 1, then, provides Wannier functions at least somewhat localized in both
space and energy.

Unless specified otherwise, we set $\gamma = 0.47714$. This is equivalent, for
lengths in \si{\angstrom} and energies in \si{\electronvolt}, to that used in
the LOSC method \cite{su2020}. In that work, $\gamma$ was chosen to minimize
delocalization error on a suite of molecules. At this value of $\gamma$, we
find that Bloch orbitals mix substantially only when their energies at a
$\bk$-point are within about \qty{2}{\electronvolt}. This natural penalty to
mixing of orbitals widely separated in energy means that dual localization
yields converged frontier orbitals irrespective of the number of high-energy
virtual states included. Details about the unit dependence of $\gamma$ may be
found in the Supplemental Material.

While $\gamma = 0.47714$ has been an effective choice in a variety of
applications in molecules as well as LOSC for materials (see
Sec.~\ref{sec:disc}), it does not satisfy a physical rule such as a variational
principle. This is a blessing and a curse; it provides a parameter than can be
tuned for the application of choice, but the character of the DLWFs (especially
the occupied DLWFs, see Fig.~\ref{fig:si.degen_cost}) depends on $\gamma$.

\subsection{Implementation details} \label{sec:dlwf.impl}
We implement the DLWF localization in a locally maintained fork 
\cite{mahler2024a} of the open-source \texttt{wannier90} code
\cite{mostofi2008, mostofi2014, pizzi2020}. Following \textcite{marzari1997} for
$\Omega$ and the previous section for $\Xi$, we compute
$\partial F / \partial U^{\bk}$ at each $\bk$-point. Given an initial guess or
disentanglement, either a conjugate gradient or steepest descent algorithm is
used to minimize $F$. Some modifications to the conjugate gradient algorithm
were required for $\Xi$, since we found that the step size that produced the
best localization was system-dependent. To account for this, we sweep a range
of step sizes in order to obtain the best minimum for each localization. For
steps when the conjugate gradient descent and parabolic line search were used,
the Polak--Ribiere coefficient \cite{polak1969} sometimes provides better
convergence than the default Fletcher--Reeves coefficient
\cite{fletcher1964, press1992}. 

Because we allow the inclusion of unoccupied orbitals, we disentangle the
highest-energy conduction bands considered for localization \cite{souza2001}.
When including virtual bands, we increase the number of bands from which DLWFs
are constructed until the orbitals of interest, usually the ones around the
Fermi energy, are converged. Convergence of high-lying virtual bands was found
to be difficult; to sidestep this issue, we implement an option allowing the
exclusion of specified bands from the cost function convergence criterion. We
consider DLWFs converged if $F$ of the included bands does not change
appreciably when higher-energy conduction bands are added. Our code may be found
at \cite{mahler2024a}; more details on the functionality added to
\texttt{wannier90} are in the Supplemental Material.

\section{Results}
We obtain Bloch functions using the PBE density functional \cite{perdew1996} in
the plane-wave basis, using optimized norm-conserving Vanderbilt
pseudopotentials \cite{hamann2013} from the PseudoDojo \cite{vansetten2018};
calculations are performed using the open-source \texttt{Quantum ESPRESSO} code
suite \cite{giannozzi2009,giannozzi2017}. For silicon and ethylene, we use a
wavefunction kinetic energy cutoff $E_{\text{cut}} = \SI{60}{\rydberg}$; for
copper, $E_{\text{cut}} = \SI{100}{\rydberg}$. In all cases, $E_{\text{cut}}$
for the density is four times that used for the wavefunctions. As mentioned
above, DLWFs are computed with a modified version of \texttt{wannier90}, with
$\gamma = 0.47714$ unless stated otherwise. We use the SCDM method
\cite{damle2015, damle2017a} as the initial guess for dual localization. DLWFs
are not always real, so the isoplots seen in
Figs.~\ref{fig:si.isos.b34w30}--\ref{fig:ethyl.isos} are of the densities
$\abs{w_{\bZ i}(\br)}^2$. Lengths are reported in \si{\angstrom} and energies
in \si{\electronvolt}. 

\subsection{Silicon} \label{sec:si}
First we consider the well-studied semiconductor silicon in the diamond lattice.
We compute the density self-consistently with a \kmt{8} $\bk$-mesh, then use a
\kmt{4} $\bk$-mesh for obtaining virtual Bloch orbitals and DLWFs. We use the
experimental lattice parameter $a = \SI{5.431}{\angstrom}$ \cite{hom1975}. The
gap of Si is small, \SI{0.71}{\electronvolt} in our PBE calculation. It is
indirect, however; the smallest direct gap in our calculation is
\SI{2.56}{\electronvolt}, at $\bk = \Gamma$. Because of this, even small values
of $\gamma$ separate silicon DLWFs into (approximately) occupied and virtual
manifolds.

\begin{table}[ht]
\caption{Energy (near) degeneracies of silicon DLWFs constructed from the
         occupied, frontier, and converged (30) bands with $\gamma = 0.47714$.
         We call two DLWFs degenerate if their average energies
         $\evt{\wt{\Ham}}{w_{\bZ i}}$ are within \SI{0.2}{\electronvolt} of
         one another. The first column is the number of degenerate DLWFs in
         each set; the other columns give $\evt{\wt{\Ham}}{w_{\bZ i}}$
         averaged over the degenerate set. Complete data may be found in the
         Supplemental Material.
        }
\label{tab:si.degen.patt}
\centering
\begin{ruledtabular}
\begin{tabular}{rddd}
    \multicolumn{1}{r}{Count} &
    \multicolumn{1}{r}{Occupied} &
    \multicolumn{1}{r}{Frontier} &
    \multicolumn{1}{r}{Converged} \\
    \colrule
    1 & -3.2290 & -3.2341 & -3.2341 \\
    1 & 0.5339 & 0.5282 & 0.5285 \\
    2 & 3.7185 & 3.7333 & 3.7388 \\
    4 &  & 11.4122 & 10.6307 \\
    2 &   &  & 15.6969 \\
\end{tabular}
\end{ruledtabular}
\end{table}

\subsubsection{Occupied states} \label{sec:si.occ}
Constructing DLWFs from only the occupied bands of silicon yields Wannier
functions which are not degenerate in energy and have different shapes. At
$\gamma = 0.47714$, the DLWF with lowest average energy
$\evt{\wt{\Ham}}{w_{\bZ i}}$ is tetrahedral. The two with highest average
energy are degenerate, each having three lobes centered along a Si--Si bond.
The shapes and degeneracy patterns are qualitatively equivalent to the four
lowest-energy DLWFs in Fig.~\ref{fig:si.degen_cost}, labeled 1--4 in
Fig.~\ref{fig:si.isos.b34w30}.

This is in contrast with with MLWFs constructed from the same set of bands, all
four of which are degenerate, bond-centered tight-binding orbitals
\cite{marzari1997}. They share the fourfold symmetry of the silicon site.
This is a tradeoff between maximally and dually localized Wannier functions:
while DLWFs retain information about the underlying energy spectrum, they do not
respect the spatial symmetries of the lattice as robustly as MLWFs do. MLWFs may
therefore be preferred in calculations of properties like the dipole moment, in
which spatial symmetry plays a major role; DLWFs are more naturally suited to
energy-dependent methods such as corrections to delocalization error (see
\ref{sec:disc} below).

\subsubsection{Frontier states} \label{sec:si.frontier}
\begin{figure}[ht]
\centering
\includegraphics[width=0.98\linewidth]{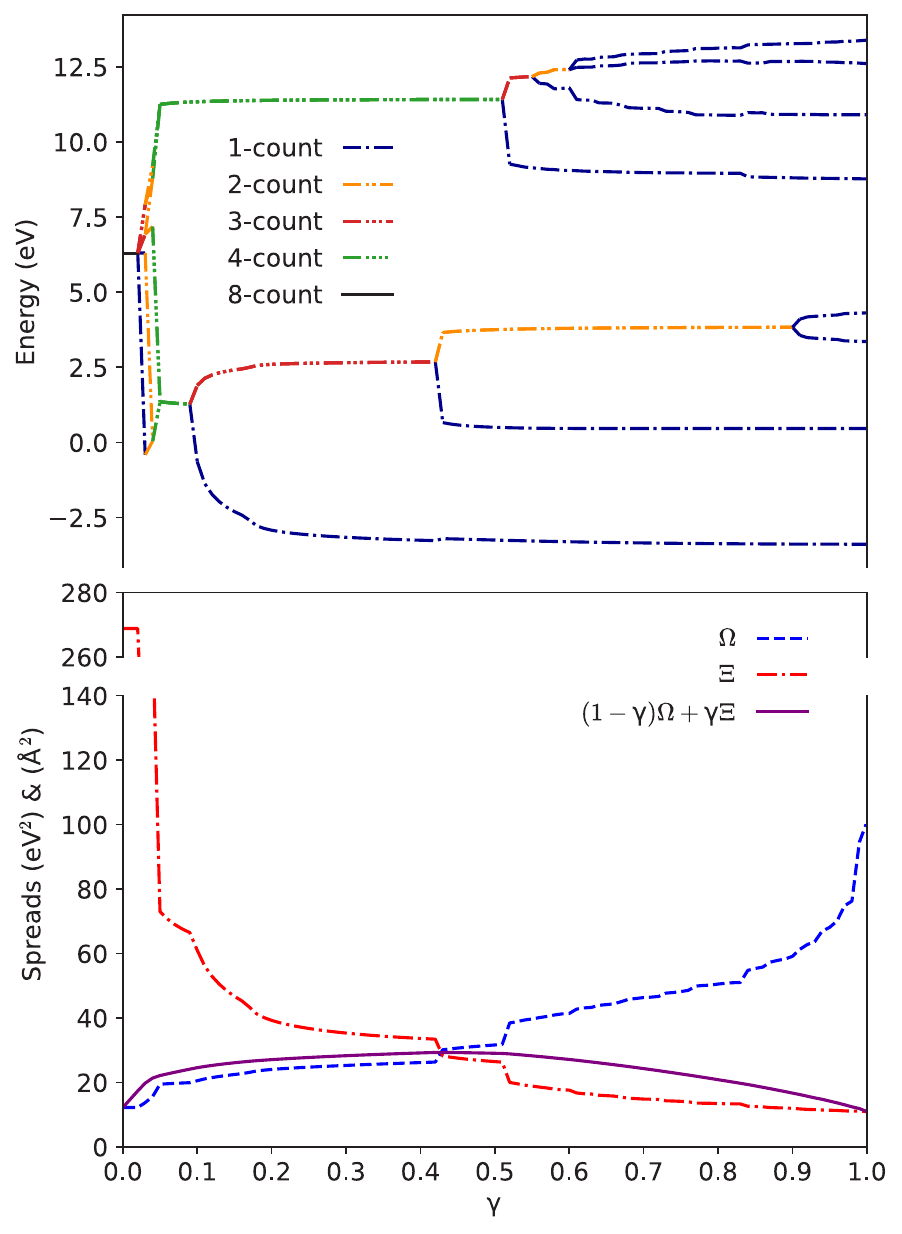}
\caption{Top panel: Degeneracy of the DLWFs in silicon as the space-energy
         mixing $\gamma$ is varied. As in Table~\ref{tab:si.degen.patt},
         we call two DLWFs degenerate if their
         average energies $\evt{\wt{\Ham}}{w_{\bZ i}}$ are within
         \qty{0.2}{\electronvolt} of one another. At $\gamma = 0$, all eight
         (maximally localized) Wannier functions are degenerate $sp^3$-like
         orbitals with occupation $\lambda_{ii} = 0.5$ and
         $\evt{\wt{\Ham}}{w_{\bZ i}} = \qty{6.3}{\electronvolt}$. At
         $\gamma = 0.05$, the DLWFs split into occupied (valence) and virtual
         (conduction) manifolds. The degeneracy of the occupied DLWFs changes
         starting at $\gamma = 0.10$, while fourfold degeneracy of the virtual
         DLWFs is not broken until $\gamma = 0.53$.
         Bottom panel: The spatial ($\Omega$, blue dashed line), energy ($\Xi$,
         red dot-dashed line), and total (purple solid line) DLWF cost functions for
         the same silicon system as $\gamma$ varies. The sharp changes in
         $\Omega$ and $\Xi$ correspond to the splitting of degenerate WFs into
         qualitatively distinct ones in the top panel. Note the discontinuity in
         the vertical axis.
        }
\label{fig:si.degen_cost}
\end{figure}

We expect the valence bands of semiconductors to yield degenerate MLWFs that
approximate bonding molecular orbitals; localizing the same number of low-lying
conduction bands with the MLWF procedure often yields degenerate functions of
antibonding character. Since the conduction states are entangled with
higher-energy bands, \textcite{souza2001} disentangled 12 bands down to 8 when
investigating silicon. Here, we use the same disentanglement with a frozen
disentanglement window at the Fermi energy, \SI{6.23}{\electronvolt}. As long
as $\gamma$ is greater than about $0.05$, the DLWFs self-organize into
(approximately)
occupied and virtual manifolds. That is, there are four DLWFs with occupations
$\lambda_{ii} = \evt{\rho}{w_{\bZ i}} \approx 1$, and four with
$\lambda_{ii} \approx 0$. (Here $\rho$ is the density matrix.) The virtual
DLWFs, which have a higher average energy $\evt{\wt{\Ham}}{w_{\bZ i}}$ than the
occupied ones, are roughly degenerate for $0.05 \leq \gamma \leq 0.53$, as can
be seen in Fig.~\ref{fig:si.degen_cost}. They resemble $\sigma^*$ antibonding
orbitals centered along different \ce{Si-Si} bonds (see DLWFs 5--8 in
Fig.~\ref{fig:si.isos.b34w30}), similar to the MLWFs constructed only from
low-energy conduction bands \cite{souza2001} or mapped to the conduction band
by automated mixing \cite{qiao2023a}. Like those in the previous section, the
occupied DLWFs do not have tight-binding character (as the MLWFs in
\cite{souza2001, qiao2023a} do), except for the narrow range around
$0.05 \leq \gamma \leq 0.10$. With a larger penalty for energy variance, a
tetrahedral DLWF with lowest average energy appears (labeled 1 in 
Fig.~\ref{fig:si.isos.b34w30}). Between $\gamma = 0.40$ and $\gamma = 0.45$, the
theefold degenerate `upper valence' DLWF submanifold splits again. It yields two
highest occupied DLWFs centered along bonds, which have the shape of
Fig.~\ref{fig:si.isos.b34w30}, DLWFs 3--4; and one reminiscent of the $sp^3$ hybrid in \cite{souza2001}, shaped like DLWF 2 in Fig.~\ref{fig:si.isos.b34w30}.
Note that such degeneracy breaking between sets of DLWFs is associated with
sharp changes in both the spatial and energy variance
(Fig.~\ref{fig:si.degen_cost}). As expected, the DLWFs' spatial variance
increases monotonically with $\gamma$, while the energy variance decreases.

The band structure interpolated from the DLWFs does not always reproduce the
DFA band structure quantitatively when the Brillouin zone is not sampled finely
enough or for high-energy conduction bands affected by disentanglement. It is
possible that a somewhat finer mesh of $\bk$-points is required for DLWFs than
for MLWFs, because dually localized Wannier functions are not guaranteed to be
exponentially localized (compare Section VI.A of \cite{marzari2012}). See
Supplemental Fig.~5 and the surrounding discussion for the effect of a larger
$\bk$-grid on band structure interpolation in silicon.

\begin{figure}[ht]
\centering
\includegraphics[width=0.98\linewidth]{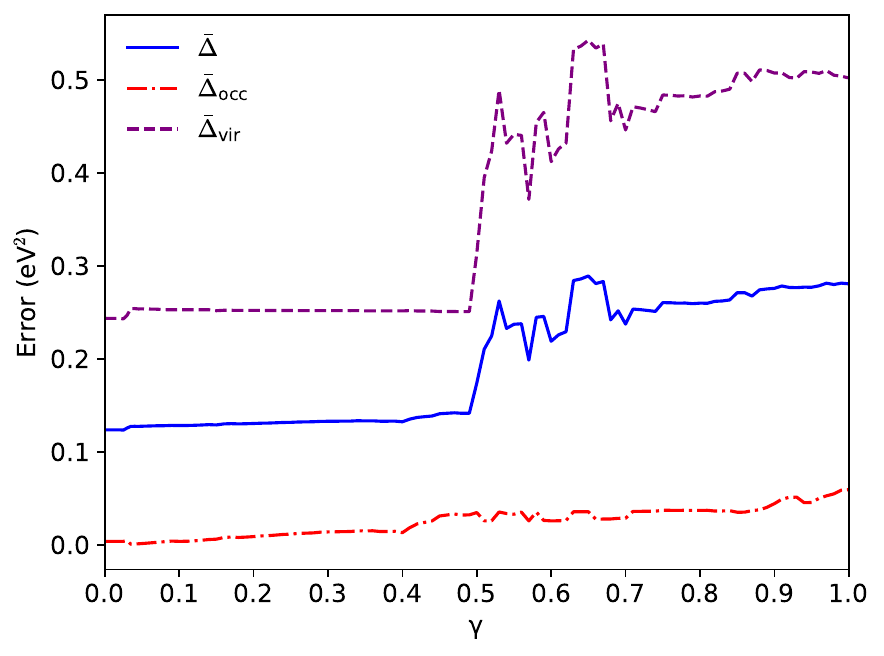}
\caption{Band interpolation error from Eq.~\eqref{eqn:interp_err} along the
         $\bk$-path used in Fig.\ \ref{fig:si.dis.b34w30}
         (L--$\Gamma$--X--K--L) for the frontier states of silicon. The solid
         blue line is the interpolation error averaged over all 8 frontier
         states; the dash-dotted red line for the 4 highest valence bands; the
         dashed purple line for the 4 lowest conduction bands.
        }
\label{fig:interp_err}
\end{figure}

In order to show how the mixing parameter $\gamma$ affects band interpolation
along a \bk-path, we calculate the difference between the interpolated
eigenvalues $\wt{\eps}_{\bk i}$ and those ($\eps_{\bk n}$) obtained before
disentanglement; the results are shown in Fig.~\ref{fig:interp_err}. (Note that
the low-energy disentangled bands $i$ of gapped systems can be reliably
identified one-to-one with original bands $n_i$.) We calculate the mean squared
error per band as
\begin{equation} \label{eqn:interp_err}
    \bar{\Delta} = 
    \frac{1}{N_w} \sum_i \left[ 
        \frac{1}{N_k} \sum_{\bk}^{\text{path}}
        \left( \eps_{\bk n_i} - \wt{\eps}_{\bk i} \right)^2
    \right];
\end{equation}
here, the sum over $\bk$ follows the band structure path, not the
Monkhorst--Pack mesh. The interpolation error in Fig.~\ref{fig:interp_err} at
$\gamma = 0$ arises primarily from the virtual orbitals; that is, from
disentanglement rather than interpolation. The noticeable upticks in
interpolation error for the valence states around $\gamma = 0.4$ and for the
conduction states around $\gamma = 0.5$ are due to groups of degenerate Wannier
functions splitting into separate states (Fig.~\ref{fig:si.degen_cost}). This
splitting of degeneracy is due to the energy cost term becoming larger than the
spatial cost term for a given DLWF configuration. For a given set of DLWFs,
then, lower-degeneracy DLWFs always have a larger spatial spread than their
higher-degeneracy counterparts. The larger spatial spread of these
lower-degeneracy DLWFs yields a poorer interpolation of the band structure
because Wannier function interpolation accuracy relies on rapid spatial decay
of the Wannier functions. A comparison of interpolated band structure plots at
some specific values of $\gamma$ is provided in the Supplemental Material.

\subsubsection{Converged frontier states} \label{sec:si.conv}
The spatial variances of the highest-energy occupied DLWFs are well converged
when 34 bands are disentangled to 30, although the average energy
$\evt{\wt{h}}{w_{\bZ i}}$ of the low-lying, essentially unoccupied DLWFs
differ between this case and that with 12 bands disentangled to 8 DLWFs
(Table \ref{tab:si.degen.patt}). We attribute this to the disentanglement
procedure. In \ref{sec:si.frontier}, the virtual DLWFs are drawn almost
entirely from the highest-energy Bloch bands, which are most impacted by
disentanglement. The lowest-energy DLWFs in this section, on the other hand,
are far from the disentanglement window because there are many more Bloch
functions, and can therefore be expected to differ somewhat. Fortunately, the
value of $\evt{\wt{h}}{w_{\bZ i}}$ does not impact calculations directly,
although (as seen in Fig. 3--4 of the Supplemental Material) the virtual bands
are better interpolated from DLWFs constructed from more Bloch orbitals. In
addition, those constructed from only 8 disentangled Bloch orbitals are more
spatially localized at the same $\gamma$; once again, we attribute this to
disentanglement. In the latter case, disentanglement directly smooths the
lowest-energy conduction bands $\ket{\wt{\psi}_{\bk n}}$, which yields more
localized DLWFs \cite{descloizeaux1964a}. With that said, the DLWFs computed
in this section are qualitatively similar to those computed only from occupied
or frontier bands, with the same pattern of degeneracies 
(Table \ref{tab:si.degen.patt}). We plot isosurfaces of the eight DLWFs of
lowest energy in Fig.~\ref{fig:si.isos.b34w30}; more data about the individual
DLWFs is in the Supplemental Material.

\begin{figure}[ht]
\centering
\includegraphics[width=\linewidth]{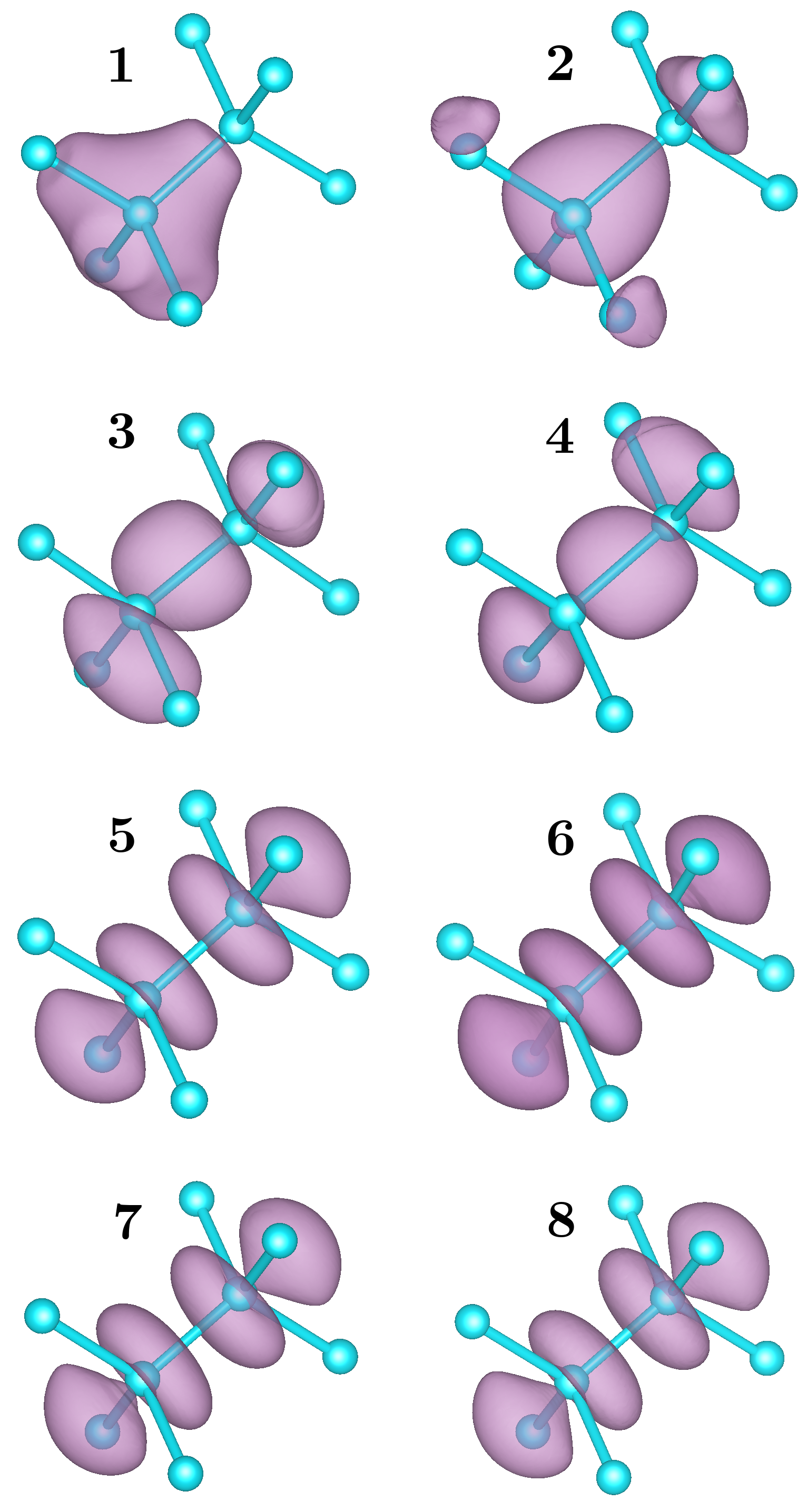}
\caption{DLWF ($\gamma = 0.47714$) density isoplots (isovalue $0.6$) for 
         silicon labeled in ascending order of $\evt{\wt{\Ham}}{w_{\bZ i}}$.
         These are obtained from 30 disentangled Bloch bands, but the
         orbitals qualitatively resemble (and have the same pattern of
         degeneracies as) those obtained from 8 disentangled frontier bands.
        }
\label{fig:si.isos.b34w30}
\end{figure}

\subsection{Copper}
\begin{figure}[ht]
\centering
\includegraphics[width=\linewidth]{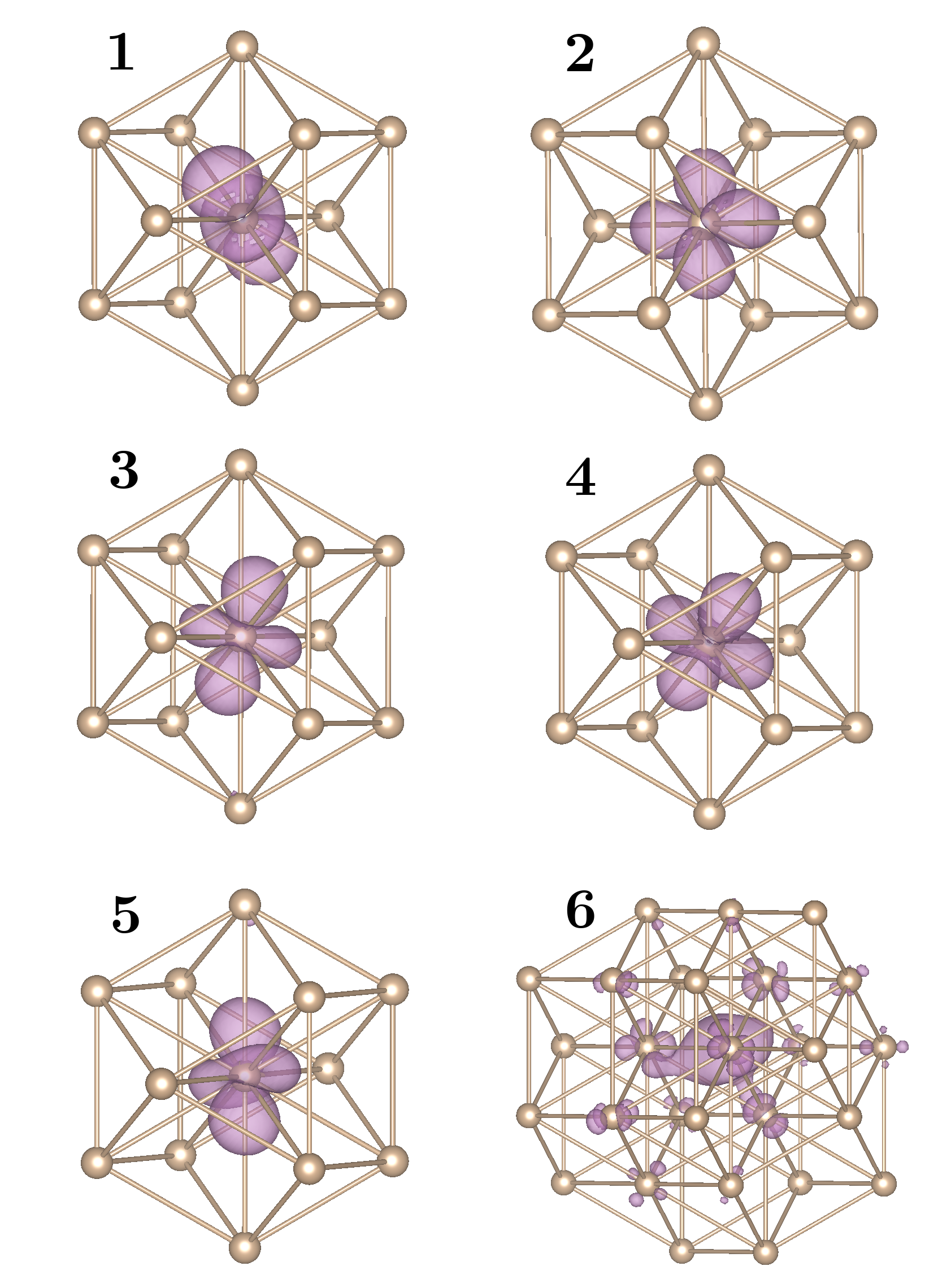}
\caption{DLWF ($\gamma = 0.47714$) density isoplots (isovalue $0.6$) of copper.
         These are the six highest-energy occupied DLWFs labeled in ascending
         order of average energy. DLWF 6 corresponds to the frontier Bloch band
         and is only partially occupied; it is also more delocalized, so more
         atoms are shown in its isoplot. (Not shown are four lower-energy DLWFs
         corresponding to $3s$ and $3p$ orbitals.)
        }
\label{fig:cu.isos}
\end{figure}

We next investigate copper in a face-centered cubic lattice, with the
experimental lattice parameter $a = \SI{3.614}{\angstrom}$ \cite{rutt2006}. For
the self-consistent calculation, we use a \kmt{16} $\bk$-mesh; for the
virtual states and localization, we use \kmt{10} $\bk$-points. Since the
unoccupied bands of copper are very steep and energy localization restricts the
mixing of bands far apart in energy, we do not require many bands above the
Fermi level. Noting that there are 9.5 electrons per spin channel, we 
disentangle 23 bands to 17 Wannier functions, with frozen disentanglement
required below \SI{40}{\electronvolt}, well above the Fermi energy
(\SI{17.04}{\electronvolt}). This results in a set of disentangled bands that
only differs from the parent DFA at the top of the energy window. When
$\gamma = 0.47714$, the fully occupied orbitals self-organize into pure $s$,
$p$, and $d$-type DLWFs (see Fig.~\ref{fig:cu.isos}). The partially occupied
DLWF associated with the frontier band has $d$-type character, but is
delocalized across multiple atoms. Information on individual DLWFs is provided
in the Supplemental Material.

\subsection{Ethylene}
\begin{figure}[ht]
\centering
\includegraphics[width=\linewidth]{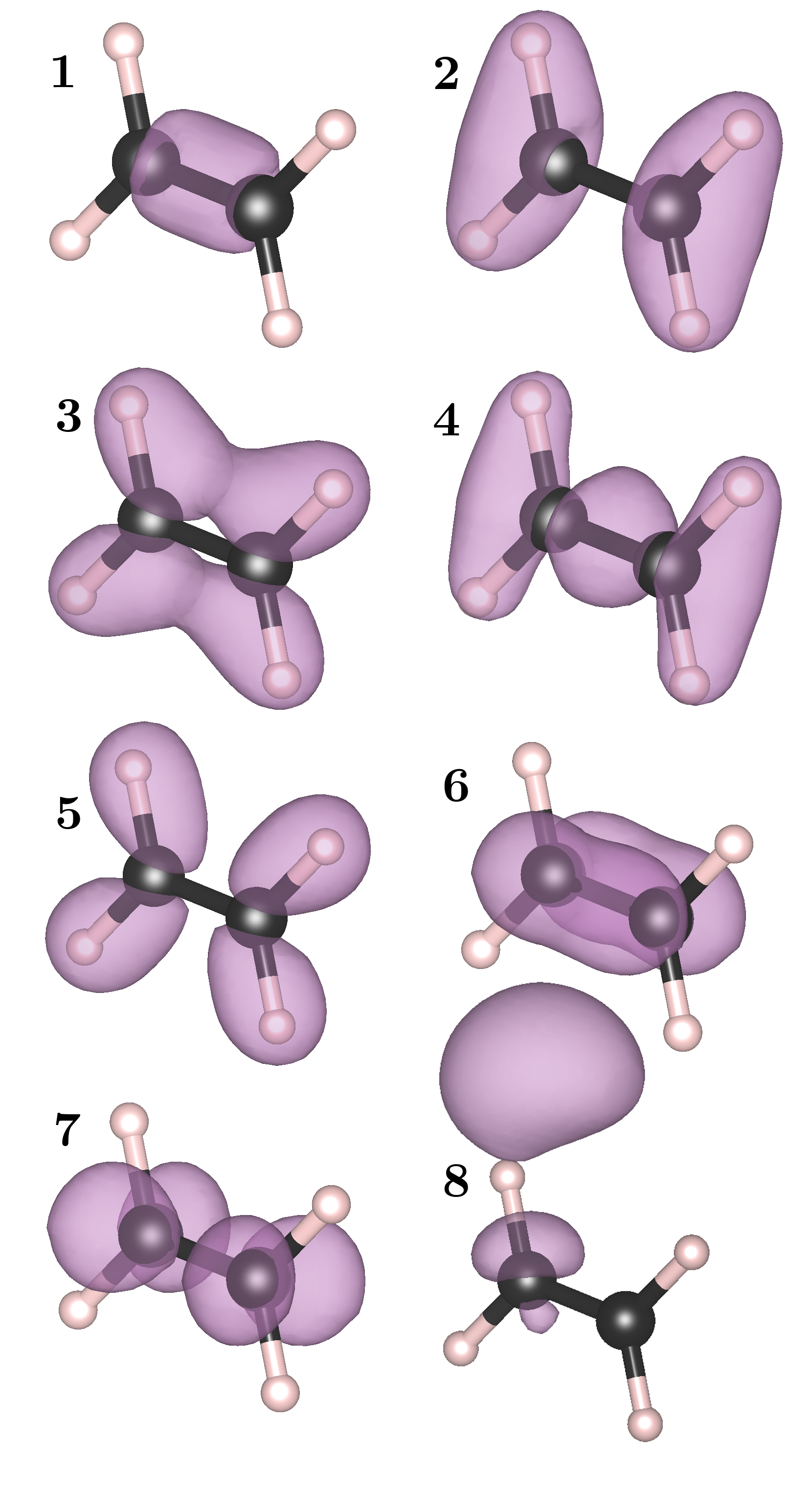}
\caption{DLWF density isoplots for ethylene using $\gamma=0.47714$. Isovalues:
         160 (DLWF 1), 60 (DLWFs 2--7), 30 (DLWF 8). The isosurface between
         DLWF 6 and DLWF 8 belongs to DLWF 8.
        }
\label{fig:ethyl.isos}
\end{figure}

To demonstrate the utility of dual localization in molecules, we simulate
ethylene with the geometry from \cite{marzari1997}. We use a \SI{10}{\angstrom}
unit cell with the molecule centered at the origin and sample only
$\bk = \Gamma$. Thus, disentanglement is unnecessary, and we construct 36 DLWFs
from 36 bands, 6 of which are occupied. Setting $\gamma = 0.47714$ produces
DLWFs closely related to molecular orbitals. The lowest-energy DLWF corresponds
to a $\sigma$ bond between the carbon atoms, and the next four resemble
combinations of \ce{C-C} and \ce{C-H} bonding orbitals. The highest-energy
occupied DLWF corresponds to a $\pi$ bonding orbital, and the lowest unoccupied
DLWF to a $\pi^*$ antibonding orbital (Fig.~\ref{fig:ethyl.isos}). The four
unoccupied DLWFs next lowest in energy are degenerate, as noted in the tabulated
DLWF information provided in the Supplemental Material, and can be characterized
as \ce{C-H} $\sigma^*$ orbitals. 

\section{Discussion} \label{sec:disc}
DLWFs have frontier orbital character in a variety of systems, from molecules
to metals. Those with occupations $\evt{\rho}{w_{\bZ i}} \approx 1$ generally
appear as bonding or atomic-like orbitals, while the low-lying unoccupied DLWFs
($\evt{\rho}{w_{\bZ i}} \approx 0$) have antibonding character. In silicon,
they separate naturally into occupied and virtual manifolds. The occupied,
atom-centered DLWF (labeled 1 in Fig.~\ref{fig:si.isos.b34w30}) is much more
localized than the other bond-centered DLWFs, in agreement with the prediction
of \textcite{kohn1973}. At $\gamma = 0.47714$, DLWFs resemble molecular
orbitals, such as $\sigma$ and $\pi$ bonding orbitals, in ethylene
(Fig.~\ref{fig:ethyl.isos}), and the $d$ manifold in copper. This shows that
DLWFs have the potential to produce chemically relevant frontier orbitals
automatically, with no need for manually enforced energy cutoffs or separation
of the valence and conduction manifolds. 

Because DLWFs offer spatially localized orbitals that describe the frontier
orbitals, they can be employed to correct delocalization error in density
functional theory. Many methods attempt to correct delocalization error, which
describes the failure of density functional approximations to give energy
piecewise linear in fractional occupation \cite{perdew1982}. They often invoke
MLWFs constructed from the valence and low-energy conduction bands separately
as localized frontier orbitals
\cite{stengel2008, borghi2014, ma2016, colonna2022} or use only valence MLWFs
\cite{wing2021}. DLWFs have fractional occupations whenever constructed from
both valence and conduction bands; $\evt{\rho}{w_{\bZ i}}$ is only
approximately, not exactly, $1$ or $0$ in gapped systems. These fractional
occupations play an important role in the LOSC method for delocalization error
correction \cite{mahler2022b, williams2024}. As DLWFs are naturally associated
with an energy level, they offer a localized charge description of frontier
orbitals for both the occupied and the unoccupied spaces, without requiring an
arbitrary limit on the number of conduction bands included. 

Analogues to DLWFs have found use in quantum chemistry, describing vibrational
states \cite{dawes2006} and many-body states for electron transfer
\cite{subotnik2009}. They are called orbitalets, in analogy to wavelets
\cite{daubechies1992}, and have proven effective in the molecular LOSC method
at improving fundamental gaps of molecules \cite{li2018, su2020, mei2020a},
polymer polarizability \cite{mei2021a}, and chemical reactivity at the frontier
energy \cite{yu2022a}. We believe that DLWFs will have similar applications in
bulk and interfacial systems.

Finally, we comment on the imaginary character sometimes present in DLWFs.
\textcite{marzari1997} conjectured that maximally localized Wannier functions
are purely real, up to a global phase. The conjecture was proven by
\textcite{brouder2007} for Wannier functions constructed by minimizing any
functional symmetric under time reversal, provided that the optimum is unique.
Our cost function $F$ obeys time-reversal symmetry, and numerical tests show
that DLWFs with $\evt{\rho}{w_{\bZ i}} \approx 1$ are real. However, we observe
substantial imaginary character in DLWFs that have an occupation numerically
distinguishable from $1$. Even though the arguments in \cite{brouder2007} do
not differentiate between the occupied and unoccupied manifolds, we conjecture
that they do not apply to Wannier functions whose occupation is far from $1$.
It is possible, however, that the imaginary character of fractionally occupied
DLWFs is because $F$ lacks a unique global minimum.

\section{Conclusion} \label{sec:concl}
We have shown that including both spatial and energy variance in the
localization cost function yields Wannier functions localized both in space and
in energy. This allows for Bloch bands widely separated in energy to be included
in their construction without mixing arbitrarily, so that DLWFs are in principle
a complete basis of localized, orthonormal, translationally symmetric orbitals.
Furthermore, the dual localization procedure naturally organizes itself along
the Hamiltonian's spectrum, yielding DLWFs with frontier-orbital character.

The ambiguity in the choice of $\gamma$, as well as the difficulty in minimizing
$F$ for high-energy conduction bands, suggests seeking an alternative---perhaps
approximate---form of DLWFs that do not require optimizing an objective
function. \textcite{damle2018} have proposed such a method, which extends the
SCDM method to entangled bands, that approximates MLWFs. It relies on the
exponential decay of the off-diagonal elements of $\rho(\br, \br')$ in space,
however, so it is unclear whether it can be adapted to the DLWF cost function.

\begin{acknowledgments}
We gratefully acknowledge helpful comments from our anonymous referees, and
financial support from the National Science Foundation (Grant No.~CHE-2154831).
A.M. was additionally supported by the Molecular Sciences Software Institute
Phase-II Software Fellowship, and J.Z.W. by the National Institutes of Health
(Grant No.~5R01GM061870).
\end{acknowledgments}

\bibliography{DLWF}

\end{document}


\setlength{\tabcolsep}{8pt}

\title{Supplemental Material for
       Wannier Functions Dually Localized in Space and Energy}
\author{Aaron Mahler}
\affiliation{Duke University, Department of Physics, Durham, NC 27708}
\author{Jacob Z. Williams}
\affiliation{Duke University, Department of Chemistry, Durham, NC 27708}
\author{Neil Qiang Su}
\affiliation{Department of Chemistry, Key Laboratory of Advanced Energy Materials Chemistry (Ministry of Education) and
Renewable Energy Conversion and Storage Center (RECAST), Nankai University, Tianjin 300071, China}
\affiliation{Duke University, Department of Chemistry, Durham, NC 27708}
\author{Weitao Yang}
\email{weitao.yang@duke.edu}
\affiliation{Duke University, Department of Chemistry, Durham, NC 27708}
\affiliation{Duke University, Department of Physics, Durham, NC 27708}

\date{\today}

\maketitle

\section{Orbital occupations}
Because we include unoccupied Bloch orbitals, we must consider orbital
occupations when considering the projector onto the occupied manifold (that is,
the one-particle density matrix):
\begin{equation}
    P_{\occ} = \frac{1}{N_k} \sum_{\bk n} f_{\bk n} \outo{\psi_{\bk n}},
\end{equation}
where $f_{\bk n} = \evt{P_{\occ}}{\psi_{\bk n}}$ is the occupation of
$\ket{\psi_{\bk n}}$ ($f_{\bk n} = 0$ or $1$ if the system is an insulator). If
the DLWFs are constructed solely from the occupied bands ($f_{\bk n} = 1$),
then also $P_{\occ} = \sum_{\bk n} \outo{w_{\bR n}}$. However, allowing valence
and conduction bands to mix means that $P_{\occ}$ is no longer diagonal in the
Wannier basis:
\begin{equation}
    P_{\occ} = 
    \sum_{\bR \bT mn} \lambda_{\bR \bT mn} \outt{w_{\bR m}}{w_{\bT n}},
\end{equation}
where $\lambda_{\bR \bT mn} = \mel{w_{\bR m}}{P_{\occ}}{w_{\bT n}}$ is the
pairwise occupation between two Wannier functions. The matrix
$\begin{pmatrix} \lambda_{\bR \bT mn} \end{pmatrix}$ is Hermitian, and while
the effective occupation $\lambda_{\bR \bR nn}$ of $\ket{w_{\bR n}}$ is not in
general an integer, it does satisfy $0 \leq \lambda_{\bR \bR nn} \leq 1$.

We can also write $P_{\occ}$ in the basis of transformed Bloch orbitals as
\begin{equation}
\begin{split}
    P_{\occ} &=
    \frac{1}{N_k^2} \sum_{\bk m} \sum_{\bq n} 
        \mel{\phi_{\bk m}}{P_{\occ}}{\phi_{\bq n}}\,
        \outt{\phi_{\bk m}}{\phi_{\bq n}} \\ &=
    \frac{1}{N_k} \sum_{\bk mn} 
        \lambda_{\bk mn} \outt{\phi_{\bk m}}{\phi_{\bq n}},
\end{split}
\end{equation}
where $\ket{\phi_{\bk n}} = \sum_m U^{\bk}_{mn} \ket{\psi_{\bk m}}$. The
pairwise occupation $\lambda_{\bk mn}$ between $\ket{\phi_{\bk m}}$ and
$\ket{\phi_{\bk n}}$ is not diagonal in the band index $n$ if valence and
conduction bands are mixed by $U^{\bk}$, although it does remain diagonal
in $\bk$; as in the Wannier basis, $0 \leq \lambda_{\bk nn} \leq 1$.

The local occupation matrices $\begin{pmatrix} \lambda_{\bk mn} \end{pmatrix}$
and $\begin{pmatrix} \lambda_{\bR \bT mn} \end{pmatrix}$ are related by discrete
Fourier transforms; and their traces, restricted to a single $\bk$ point (or
unit cell $\bR$), give the number of electrons below the Fermi energy at $\bk$
(in the unit cell $\bR$). That is,
\begin{equation}
    \Tr_{\bk}{\begin{pmatrix} \lambda_{\bk mn} \end{pmatrix}} =
    \sum_{n} \lambda_{\bk nn} = N_{\occ}(\bk),
\end{equation}
and
\begin{equation}
    \Tr_{\bR}{\begin{pmatrix} \lambda_{\bR \bT mn} \end{pmatrix}} =
    \sum_{n} \lambda_{\bR \bR nn} = 
    \frac{1}{N_k} \sum_{\bk} N_{\occ}(\bk).
\end{equation}

\section{Gauge-invariant energy cost with disentanglement}
The projector $\wt{P}$ onto the subspace spanned by the disentangled bands
$\ket{\wt{\psi}_{\bk i}}$ can be written in either the disentangled Bloch or
Wannier basis:
\begin{equation}
    \wt{P} =
    \sum_{\bk i} \outo{\wt{\psi}_{\bk i}} =
    \sum_{\bR i} \outo{w_{\bR i}}.
\end{equation}
Then we can write the gauge-invariant part of the energy cost $\Xi$ as
\begin{equation}
\begin{split}
    \Xi_{\text{I}} &=
    \sum_i \left[ 
        \evt{\wt{h}^2}{w_{\bZ i}} - 
        \sum_{\bR j} \abs{\mel{w_{\bR j}}{\wt{h}}{w_{\bZ i}}}^2 
    \right] \\ &=
    \sum_i \left[ 
        \evt{\wt{h}^2}{w_{\bZ i}} -
        \sum_{\bR j} 
            \mel{w_{\bZ i}}{\wt{h}}{w_{\bR j}} 
            \mel{w_{\bR j}}{\wt{h}}{w_{\bZ i}}
    \right] \\ &=
    \sum_i \left[
        \evt{\wt{h}^2}{w_{\bZ i}} -
        \evt{\wt{h} \left( \sum_{\bR j} \outo{w_{\bR j}} \right) \wt{h}}{w_{\bZ i}}
    \right] \\ &=
    \sum_i \left[
        \evt{\wt{h}^2}{w_{\bZ i}} - \evt{\wt{h}\wt{P}\wt{h}}{w_{\bZ i}}
    \right] =
    \sum_i \evt{\wt{h} I \wt{h} - \wt{h} \wt{P} \wt{h}}{w_{\bZ i}} \\ &=
    \sum_i \evt{\wt{h}^2}{w_{\bZ i}} - 
        \sum_i \evt{\wt{h}\wt{P}\wt{h}}{w_{\bZ i}} = \\ &=
    \Tr_c{[\wt{P}\wt{h}^2]} - \Tr_c{[\wt{P}\wt{h}\wt{P}\wt{h}]}.
\end{split}
\end{equation}
Noting that
\begin{equation}
\begin{split}
    \wt{P} \wt{h} &= 
    \left( 
        \sum_{\bk i} \outo{\wt{\psi}_{\bk i}}
    \right) \left( 
        \sum_{\bk' i'} \wt{\eps}_{\bk' i'} \outo{\wt{\psi}_{\bk' i'}}
    \right) \\ &=
    \sum_{\bk\bk' ii'} \wt{\eps}_{\bk' i'} 
        \ket{\wt{\psi}_{\bk i}} 
        \inn{\wt{\psi}_{\bk i}}{\wt{\psi}_{\bk' i'}} 
        \bra{\wt{\psi}_{\bk' i'}} \\ &=
    \sum_{\bk i} \wt{\eps}_{\bk i} \outo{\wt{\psi}_{\bk i}} \\ &=
    \wt{h},
\end{split}
\end{equation}
whence $\wt{P}\wt{h}^2 = \wt{P}\wt{h}\wt{P}\wt{h} = \wt{h}^2$, we obtain that
\begin{equation}
    \Xi_{\text{I}} = 
    \Tr_c{[\wt{P}\wt{h}^2]} - \Tr_c{[\wt{P}\wt{h}\wt{P}\wt{h}]} = 0.
\end{equation}

\section{Appendix of Tables}
This section shows the spatial center ($\evo{r_x}, \evo{r_y}, \evo{r_z}$),
spatial variance $\evo{\Delta r^2}$, average energy $\evo{h}$, energy variance
$\evo{\Delta h^2}$, and occupation $\lambda_{\bZ ii} = \evt{\rho}{w_{\bZ i}}$
of the individual DLWFs presented in the Results section of the main text. The
index is ordered from lowest to highest energy, starting at 1. Thus, the index
in Table \ref{tab:si_b34w30} (Table \ref{tab:copper}, Table~\ref{tab:ethylene})
below corresponds to the DLWF density isoplot labels in Fig.~4 (Fig.~5, Fig.~6)
of the main text.

\subsection{Silicon} \label{sec:si.app}
\begin{longtable}{@{}cccccccc@{}}
\caption{Spatial and energy information per DLWF when only using the valence states of silicon. }
\label{tab:si_b4w4} \\
        \toprule
        index & $\langle r_x \rangle$ & $\langle r_y \rangle$ & $\langle r_z \rangle$ & $\langle \Delta r^2 \rangle$ & $\langle h \rangle$ & $\langle \Delta h^2 \rangle$ & $\lambda_{\bZ ii}$ \\
        \colrule
        1 & 0.6280 & 0.7208 & 0.7298 & 2.346988 & -3.228980 & 1.700047 & 1.000000 \\
        2 & -1.1404 & -0.8386 & -0.2173 & 4.867715 & 0.533916 & 3.614710 & 1.000000 \\
        3 & 0.9968 & 0.0737 & 1.6037 & 4.481121 & 3.718522 & 1.627641 & 1.000000 \\
        4 & -0.2461 & 0.0736 & 0.3609 & 4.481166 & 3.718535 & 1.627611 & 1.000000 \\
        \botrule
\end{longtable}

\begin{longtable}{@{}cccccccc@{}}
\caption{Spatial and energy information per DLWF when using 12 bands disentangled to produce 8 DLWFs. }
\label{tab:si_b12w8} \\
        \toprule
        index & $\langle r_x \rangle$ & $\langle r_y \rangle$ & $\langle r_z \rangle$ & $\langle \Delta r^2 \rangle$ & $\langle h \rangle$ & $\langle \Delta h^2 \rangle$ & $\lambda_{\bZ ii}$ \\
        \colrule
        1 & 0.6393 & 0.7274 & 0.7274 & 2.358287 & -3.234109 & 1.681093 & 0.999994 \\
        2 & -0.5300 & -0.2339 & -0.2340 & 4.840094 & 0.528172 & 3.587587 & 0.999599 \\
        3 & -1.4288 & -1.1217 & 0.3475 & 4.369317 & 3.733272 & 1.664589 & 0.998669 \\
        4 & 1.2869 & 0.3475 & 1.5940 & 4.369209 & 3.733309 & 1.664698 & 0.998666 \\
        5 & 0.0548 & 0.0466 & 0.0468 & 3.821754 & 11.392893 & 4.591103 & 0.000972 \\
        6 & -0.0529 & 1.4009 & 1.4007 & 3.817564 & 11.396780 & 4.589852 & 0.000944 \\
        7 & -1.3560 & -0.0132 & 4.0654 & 3.822250 & 11.428582 & 4.551941 & 0.000580 \\
        8 & -1.3552 & 1.3505 & 2.7017 & 3.819881 & 11.430745 & 4.553351 & 0.000577 \\
        \botrule
\end{longtable}

\begin{longtable}{@{}cccccccc@{}}
        \caption{ Spatial and energy information per DLWF for silicon using 34 bands disentangled 
        to 30 bands. }
        \label{tab:si_b34w30} \\
        \toprule
        index & $\langle r_x \rangle$ & $\langle r_y \rangle$ & $\langle r_z \rangle$ & $\langle \Delta r^2 \rangle$ & $\langle h \rangle$ & $\langle \Delta h^2 \rangle$ & $\lambda_{\bZ ii}$ \\
        \colrule
        1 & -0.7273 & -0.6394 & -0.7274 & 2.357579 & -3.234141 & 1.681126 & 0.999992 \\
        2 & 0.2355 & 0.5311 & 0.2360 & 4.824702 & 0.528461 & 3.594108 & 0.999453 \\
        3 & -0.3476 & -1.2893 & -1.5917 & 4.296241 & 3.738779 & 1.693098 & 0.997848 \\
        4 & -1.5941 & -1.2878 & -0.3466 & 4.294985 & 3.738866 & 1.692406 & 0.997860 \\
        5 & -0.0505 & 2.6608 & 2.7073 & 4.845865 & 10.548782 & 4.084945 & 0.001534 \\
        6 & 1.2678 & 0.0730 & 1.2995 & 4.853594 & 10.569209 & 4.085532 & 0.001481 \\
        7 & 1.4140 & -1.3094 & -2.7687 & 4.916316 & 10.693759 & 4.064643 & 0.000875 \\
        8 & 0.0238 & 1.3053 & 1.3575 & 4.907699 & 10.711160 & 4.088860 & 0.000869 \\
        9 & -0.0182 & 0.0123 & 0.0392 & 8.239568 & 15.653925 & 3.218761 & 0.000047 \\
        10 & -0.8270 & -0.8582 & -1.3807 & 8.298693 & 15.739891 & 2.969253 & 0.000037 \\
        11 & 0.1125 & -1.2840 & -1.3794 & 8.389856 & 18.404274 & 1.924312 & 0.000001 \\
        12 & -0.7443 & -0.7126 & 0.0674 & 8.252689 & 18.922132 & 1.841812 & 0.000002 \\
        13 & 1.0858 & -0.0787 & 1.4867 & 8.336248 & 19.346286 & 2.670695 & 0.000001 \\
        14 & -0.1180 & -1.1155 & 1.2004 & 9.336795 & 21.777131 & 3.725877 & 0.000000 \\
        15 & 0.0191 & -1.2920 & -1.3532 & 8.073594 & 25.390255 & 2.437707 & 0.000000 \\
        16 & 0.0175 & 0.0323 & -0.0157 & 8.196600 & 25.395202 & 2.166016 & 0.000000 \\
        17 & 1.3469 & 0.2140 & 1.1390 & 8.188723 & 25.429163 & 2.396005 & 0.000000 \\
        18 & -1.4098 & 0.0256 & -1.3264 & 7.304936 & 25.916725 & 3.080916 & 0.000000 \\
        19 & 1.3965 & 0.0338 & -1.3458 & 9.008920 & 26.269281 & 2.215705 & 0.000000 \\
        20 & 0.3826 & 0.3316 & 0.4795 & 10.802935 & 29.338486 & 3.072971 & 0.000000 \\
        21 & -0.0229 & -2.7464 & -2.7040 & 8.376918 & 31.775110 & 3.252854 & 0.000000 \\
        22 & -1.3644 & -2.7038 & 1.3521 & 8.258431 & 31.791921 & 3.342923 & 0.000000 \\
        23 & -0.0453 & -1.3281 & -1.3151 & 8.891537 & 32.291000 & 2.209553 & 0.000000 \\
        24 & -0.0594 & -1.3083 & 1.3658 & 9.803869 & 34.805345 & 3.125115 & 0.000000 \\
        25 & 1.1648 & -1.2274 & 0.1947 & 9.635747 & 35.629707 & 3.195436 & 0.000000 \\
        26 & -2.6160 & -0.0473 & 0.0126 & 9.687295 & 35.795476 & 3.419050 & 0.000000 \\
        27 & 2.5350 & -0.1019 & -2.5121 & 9.086322 & 37.532297 & 3.298730 & 0.000000 \\
        28 & 1.3284 & -0.0261 & -1.3222 & 8.512641 & 39.136116 & 3.739278 & 0.000000 \\
        29 & 0.0396 & 2.7316 & -0.0278 & 9.142814 & 39.527945 & 3.099532 & 0.000000 \\
        30 & 0.0056 & 1.3209 & 1.3000 & 9.366045 & 40.314612 & 2.403746 & 0.000000 \\
        \botrule
\end{longtable}

\subsection{Copper} \label{sec:cu.app}
\begin{table}[H]
\caption{\label{tab:copper} Spatial and energy information per DLWF for copper. }
    \begin{center}
    \begin{tabular}{@{}cccccccc@{}}
        \toprule
        index & $\langle r_x \rangle$ & $\langle r_y \rangle$ & $\langle r_z \rangle$ & $\langle \Delta r^2 \rangle$ & $\langle h \rangle$ & $\langle \Delta h^2 \rangle$ & $\lambda_{\bZ ii}$ \\
        \colrule
        01 & -0.0000 & 0.0000 & -0.0000 & 0.155525 & -95.528687 & 0.000018 & 1.000000 \\
        02 & 0.0000 & 0.0000 & 0.0000 & 0.186959 & -52.928957 & 0.000814 & 1.000000 \\
        03 & 0.0000 & -0.0000 & 0.0000 & 0.186943 & -52.928955 & 0.000811 & 1.000000 \\
        04 & -0.0000 & -0.0000 & -0.0000 & 0.186958 & -52.928954 & 0.000811 & 1.000000 \\
        05 & -0.1631 & -0.0051 & 0.0045 & 1.326589 & 13.553628 & 3.061090 & 0.996669 \\
        06 & -0.0038 & -0.0046 & -0.0073 & 0.705379 & 14.072693 & 1.235276 & 0.996415 \\
        07 & -0.0063 & 0.0014 & -0.0177 & 0.793226 & 14.078952 & 1.528006 & 0.996099 \\
        08 & -0.0068 & -0.0050 & 0.0046 & 0.709318 & 14.085334 & 1.248489 & 0.996337 \\
        09 & 0.0096 & 0.0045 & -0.0013 & 0.803549 & 14.121215 & 1.528358 & 0.996831 \\
        10 & 1.0363 & 0.9041 & -0.8420 & 4.926801 & 16.988522 & 7.848977 & 0.516482 \\
        11 & 0.6418 & -1.0563 & -0.3213 & 7.008280 & 26.141562 & 16.535585 & 0.000001 \\
        12 & -0.4665 & -0.0404 & 1.4063 & 7.196124 & 32.475400 & 19.176457 & 0.000000 \\
        13 & -0.4218 & 1.0018 & 1.0964 & 7.318156 & 32.739401 & 20.308781 & 0.000000 \\
        14 & 0.0123 & 0.0419 & -0.0969 & 12.324220 & 39.582046 & 8.504523 & 0.000000 \\
        15 & 1.1162 & -0.7577 & -1.2117 & 13.706412 & 45.246294 & 12.532017 & 0.001148 \\
        16 & 1.1295 & -1.5696 & -1.3568 & 11.206204 & 51.485678 & 15.577288 & 0.000009 \\
        17 & 1.1818 & -0.9269 & 0.0903 & 8.954952 & 55.296748 & 12.405831 & 0.000009 \\
        \botrule
    \end{tabular}
    \end{center}
\end{table}
	
\pagebreak
\subsection{Ethylene} \label{sec:et.app}

\begin{longtable}{@{}cccccccc@{}}
    \caption{Spatial and energy information per DLWF for ethylene, truncated to the lowest 20 states. }
    \label{tab:ethylene} \\
    \toprule
    index & $\langle r_x \rangle$ & $\langle r_y \rangle$ & $\langle r_z \rangle$ & $\langle \Delta r^2 \rangle$ & $\langle h \rangle$ & $\langle \Delta h^2 \rangle$ & $\lambda_{\bZ ii}$ \\
    \colrule
    1 & -0.0000 & -0.0000 & 0.0000 & 0.983945 & -18.612524 & 0.000174 & 1.000000 \\
    2 & 0.0001 & 0.0000 & 0.0000 & 1.784198 & -13.889465 & 0.000596 & 0.999997 \\
    3 & 0.0002 & -0.0000 & -0.0001 & 1.633431 & -11.218850 & 0.003959 & 0.999974 \\
    4 & -0.0004 & 0.0000 & 0.0000 & 1.497676 & -9.940055 & 0.005812 & 0.999947 \\
    5 & -0.0010 & -0.0000 & -0.0000 & 2.015195 & -8.185370 & 0.007015 & 0.999939 \\
    6 & 0.0000 & 0.0000 & -0.0002 & 1.329210 & -6.559715 & 0.010217 & 0.999856 \\
    7 & 0.0007 & 0.0000 & 0.0014 & 2.009921 & -0.717311 & 0.135739 & 0.000000 \\
    8 & -1.1695 & 2.1974 & -0.1999 & 3.337830 & 1.292702 & 1.286921 & 0.000025 \\
    9 & -1.1696 & -2.1974 & -0.2001 & 3.337846 & 1.292738 & 1.286973 & 0.000025 \\
    10 & 2.2760 & -1.8212 & -0.0673 & 3.315293 & 1.349986 & 1.214562 & 0.000036 \\
    11 & 2.2761 & 1.8212 & -0.0673 & 3.315328 & 1.350045 & 1.214551 & 0.000036 \\
    12 & -3.3293 & 0.0001 & -0.0157 & 3.786638 & 1.493888 & 1.128354 & 0.000011 \\
    13 & -0.0724 & 0.0004 & -2.7003 & 3.546867 & 1.955814 & 1.377605 & 0.000052 \\
    14 & -0.0642 & 1.5299 & 2.4554 & 4.019978 & 2.136463 & 1.134380 & 0.000044 \\
    15 & -0.0651 & -1.5302 & 2.4552 & 4.020222 & 2.136698 & 1.134326 & 0.000044 \\
    16 & -0.1124 & 3.2622 & -3.9195 & 4.562633 & 2.820583 & 1.205123 & 0.000002 \\
    17 & -0.1124 & -3.2610 & -3.9198 & 4.562800 & 2.820617 & 1.205088 & 0.000002 \\
    18 & -4.4466 & -3.2543 & -0.1016 & 4.613289 & 2.977994 & 1.161815 & 0.000002 \\
    19 & -4.4465 & 3.2544 & -0.1015 & 4.613393 & 2.978025 & 1.161790 & 0.000002 \\
    20 & -3.4117 & 4.9992 & 3.3338 & 4.192816 & 2.989893 & 1.415940 & 0.000000 \\
    \botrule
\end{longtable}

\pagebreak
\section{Band Structures}
Here we show the disentangled band structures used for the systems reported in
the main text. Observe that the disentangled band structure agrees almost exactly
with the full band structure near the Fermi energy. We found that DLWFs with a
mixing parameter of $\gamma=0.01$ gave the best interpolation, which was
slightly better than using $\gamma=0.0$ (MLWF) and noticeably better than the
value of $\gamma=0.47714$ used in the isosurface plots. See Supplemental
Sec.~\ref{sec:interp_err} for details.

\begin{figure}[!htp]
\centering
\includegraphics[scale=1.0]{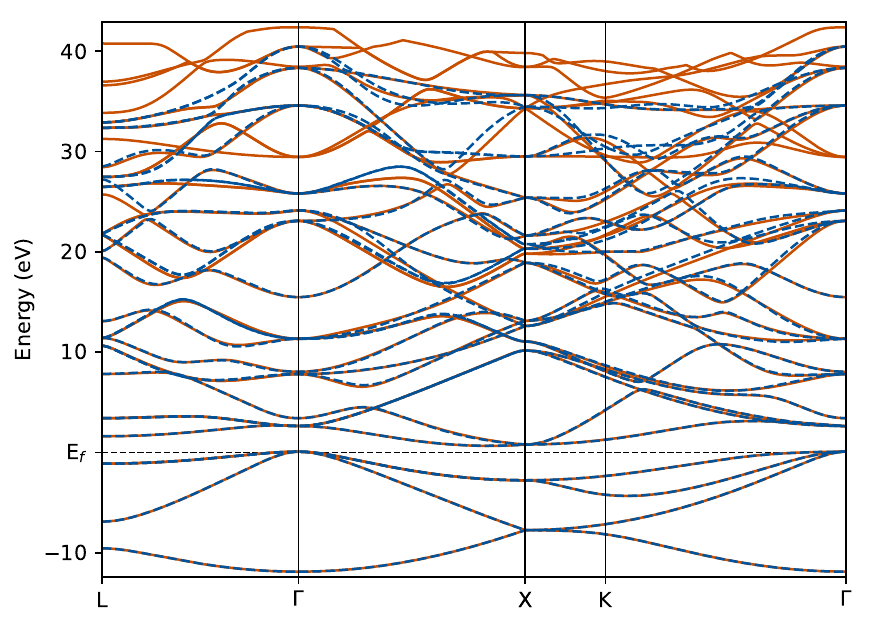}
\caption{Disentangled band structure of silicon used for the isosurface plots in
the main text. The solid orange lines are 34 bands from a non-self-consistent
calculation in {\tt Quantum ESPRESSO} with the zero point of the $y$-axis set at
the Fermi energy. The dashed blue lines are 30 disentangled bands interpolated
along the path through the Brillouin zone using DLWFs with a mixing parameter of
$\gamma=0.01$. }
\label{fig:si.evg}
\end{figure}

\begin{figure}[!htp]
\centering
\includegraphics[scale=1.0]{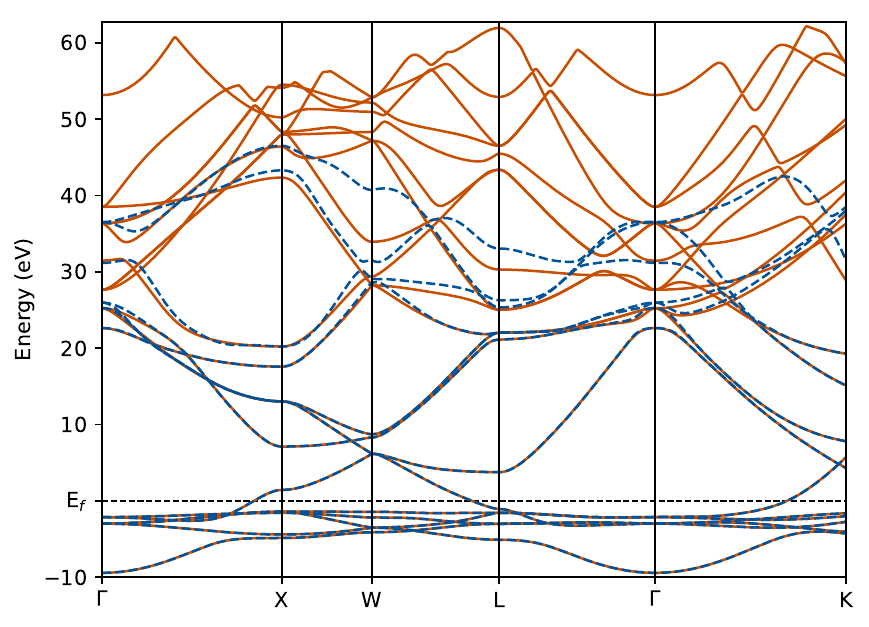}
\caption{Disentangled band structure of copper used for the isosurface plots in
the main text. The solid orange lines are 23 bands from a non-SCF calculation in
{\tt Quantum ESPRESSO}, and the energy axis is zeroed at the Fermi energy. The
dashed blue lines are 17 disentangled bands interpolated along the path through
the Brillouin zone using DLWFs with a mixing parameter of $\gamma=0.01$.}
\label{fig:cu.evg}
\end{figure}

\pagebreak
\subsection{Interpolated Band Structure Errors in Silicon} \label{sec:interp_err}
\begin{figure}[!htp]
\centering
\subfloat[$\gamma=0.000$]{\label{fig:si.b12w8a}%
  \includegraphics[width=0.48\linewidth]{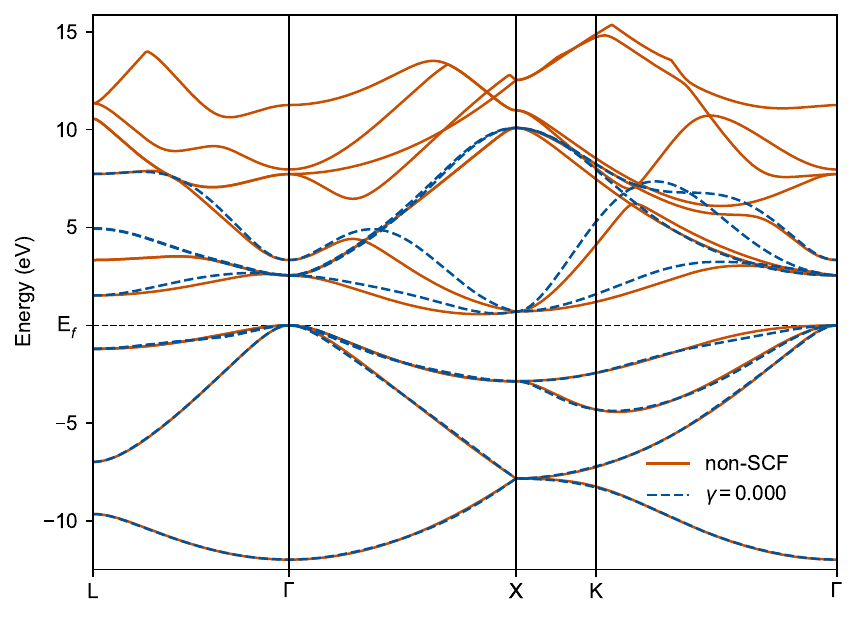}%
} 
\subfloat[$\gamma=0.001$]{\label{fig:si.b12w8b}%
  \includegraphics[width=0.48\linewidth]{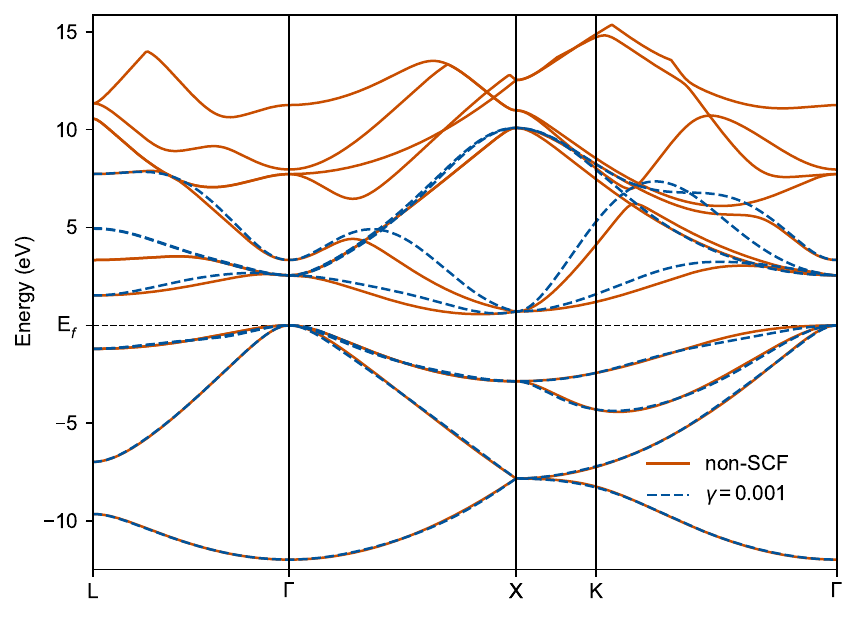}%
} \\
\subfloat[$\gamma=0.100$]{\label{fig:si.b12w8c}%
  \includegraphics[width=0.48\linewidth]{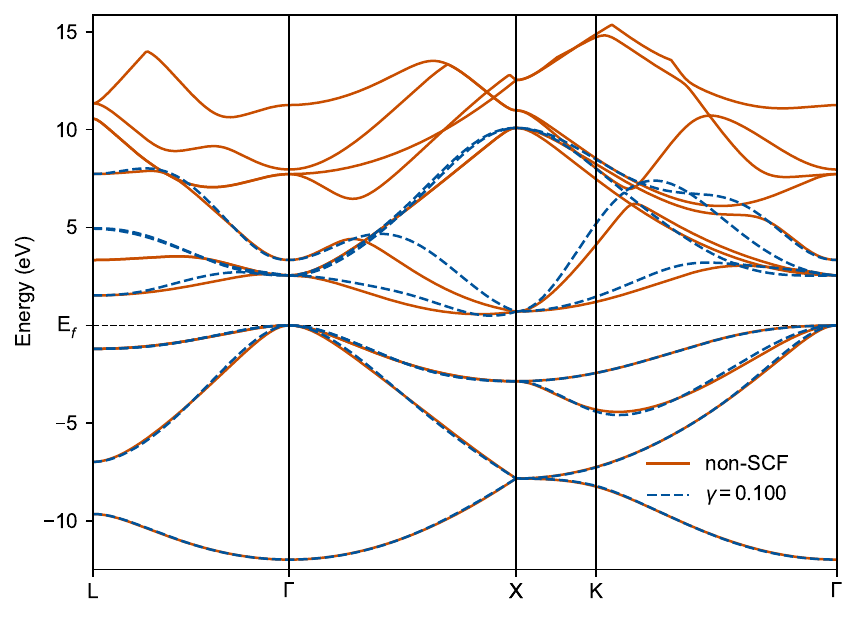}%
}
\subfloat[$\gamma=0.200$]{\label{fig:si.b12w8d}%
  \includegraphics[width=0.48\linewidth]{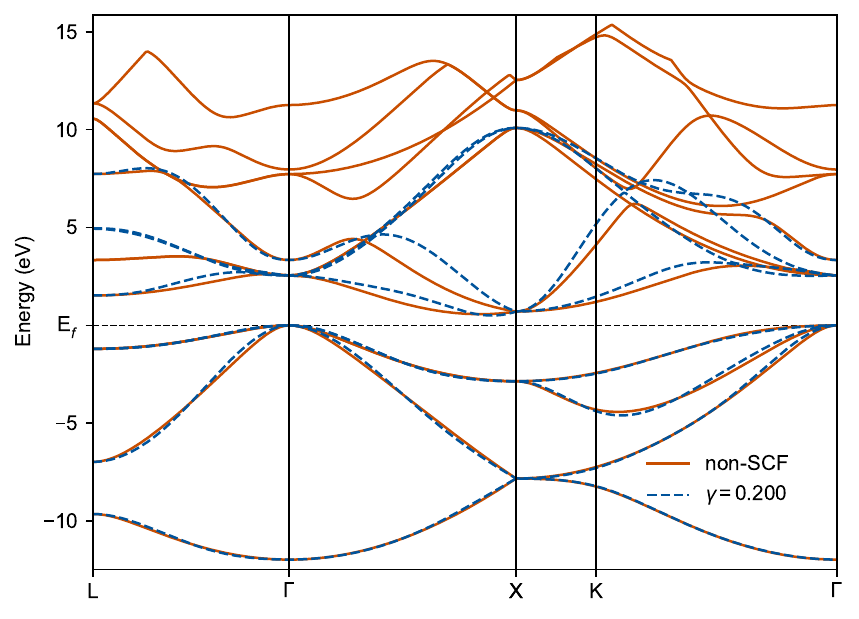}%
} \\
\subfloat[$\gamma=0.400$]{\label{fig:si.b12w8e}%
  \includegraphics[width=0.48\linewidth]{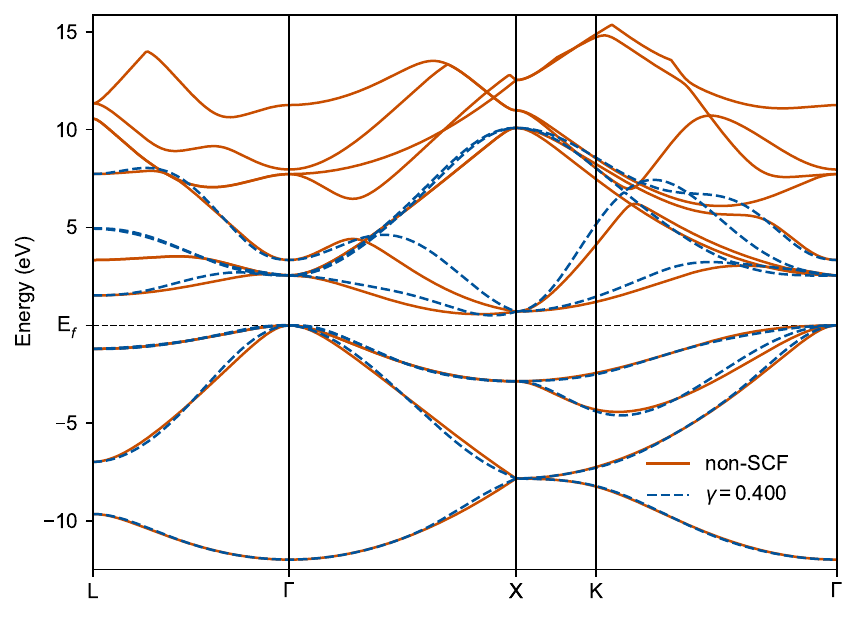}%
}
\subfloat[$\gamma=0.600$]{\label{fig:si.b12w8f}%
  \includegraphics[width=0.48\linewidth]{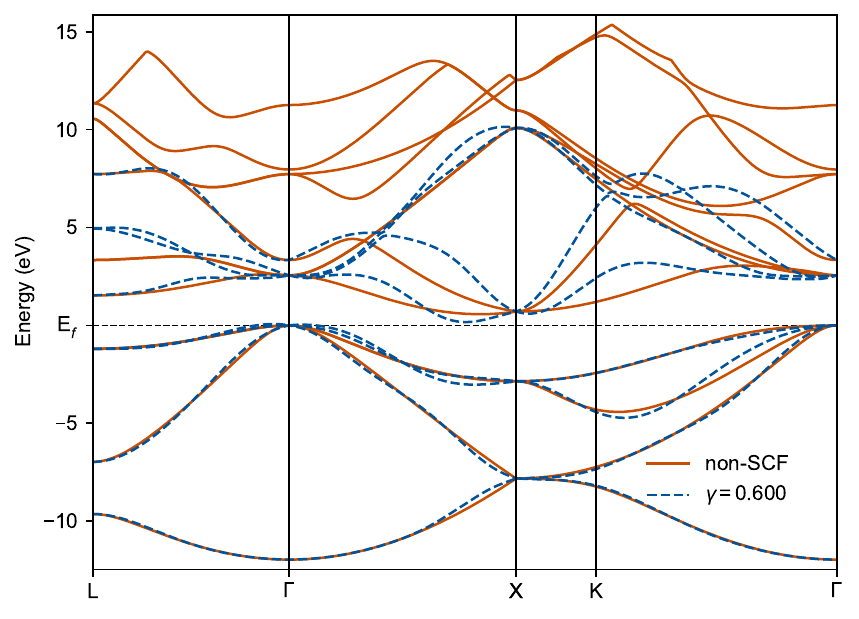}%
} \\
\caption{
Interpolated band structures for the frontier states of silicon (12 bands and 8 Wannier functions) for varying values of $\gamma$.
}
\end{figure}

\begin{figure}[!htp]
\subfloat[$\gamma=0.0$]{\label{fig:si.b34w30a}%
  \includegraphics[width=0.48\linewidth]{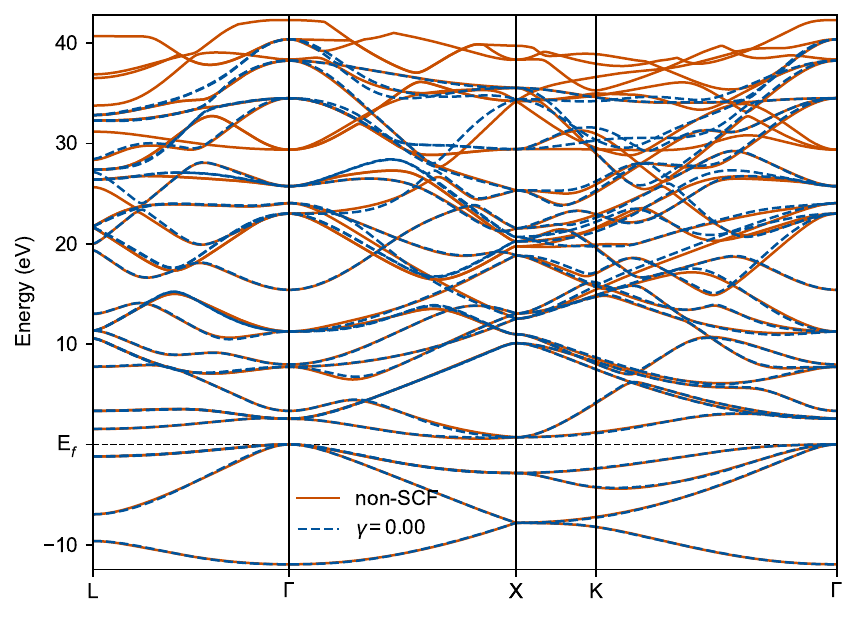}%
} 
\subfloat[$\gamma=0.01$]{\label{fig:si.b34w30b}%
  \includegraphics[width=0.48\linewidth]{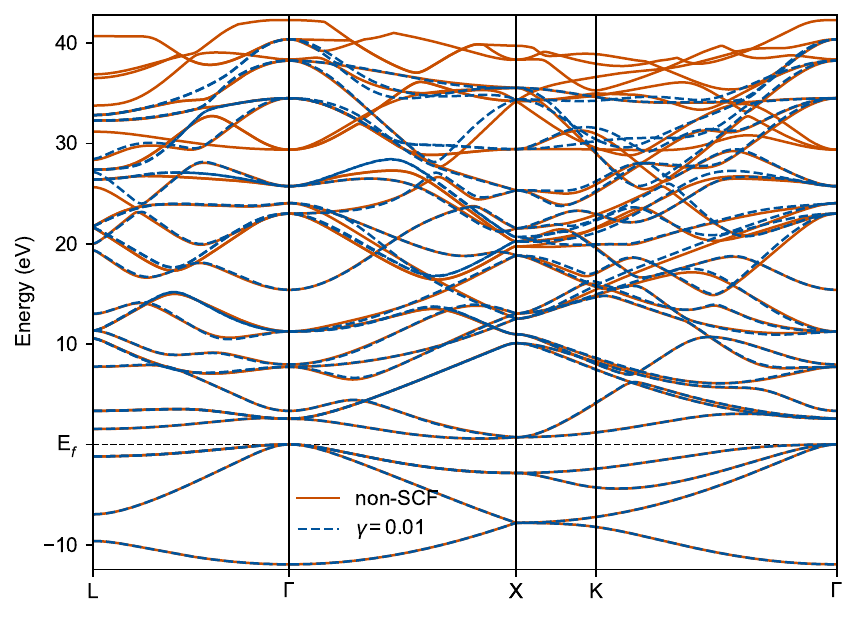}%
} \\
\subfloat[$\gamma=0.1$]{\label{fig:si.b34w30c}%
  \includegraphics[width=0.48\linewidth]{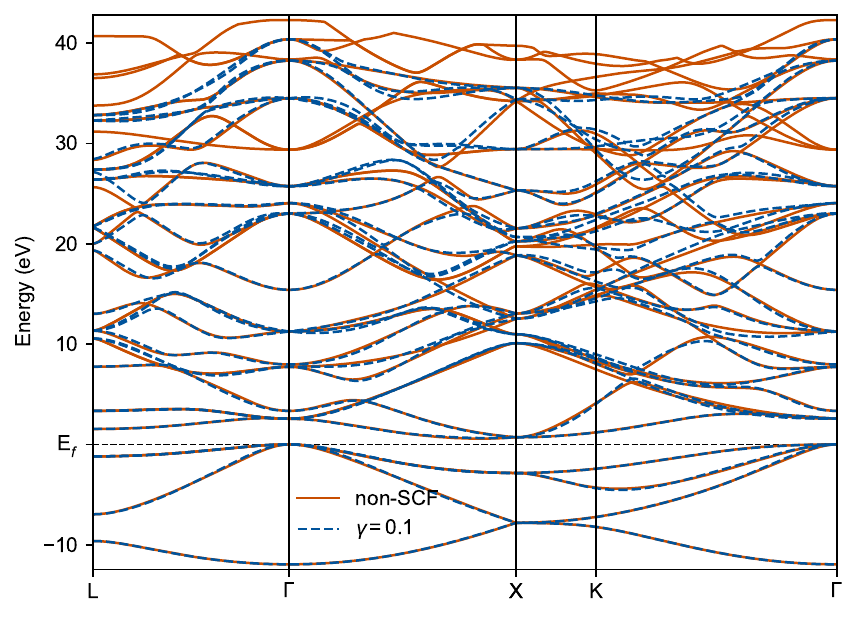}%
}
\subfloat[$\gamma=0.2$]{\label{fig:si.b34w30d}%
  \includegraphics[width=0.48\linewidth]{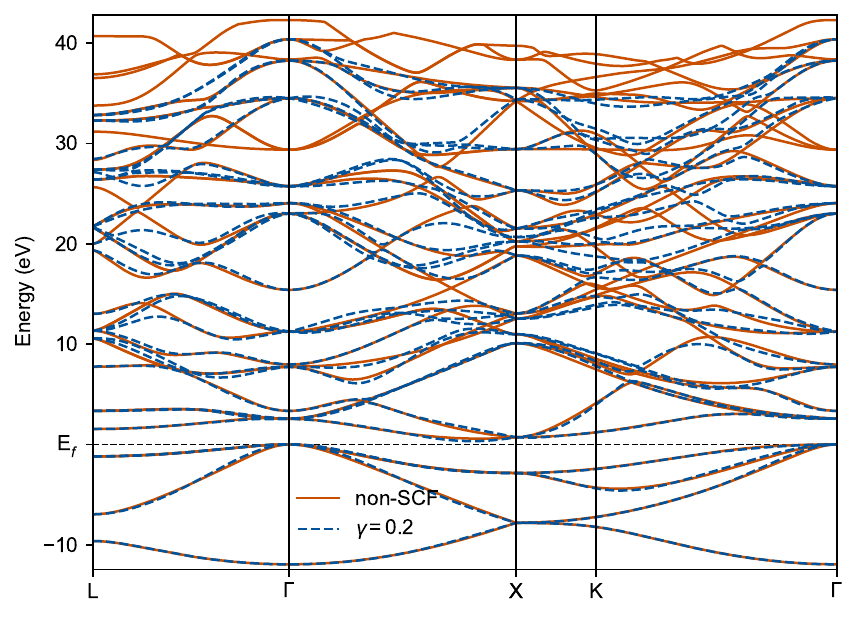}%
} \\
\subfloat[$\gamma=0.47714$]{\label{fig:si.b34w30e}%
  \includegraphics[width=0.48\linewidth]{b34w30_se0.47714.pdf}%
} \\
\caption{
Interpolated band structures for the converged states of silicon (34 bands and 30 Wannier functions) for varying values of $\gamma$. 
}
\end{figure}

Here we show some interpolated band structure plots for specific values of
$\gamma$ in the case of silicon. The plots shown in
Figs.~\ref{fig:si.b12w8a}--\ref{fig:si.b12w8f} are for 12 bands disentangled to
produce 8 Wannier functions, the average interpolation error which is reported
in the main text. For the case of 34 bands disentangled down to 30, the average
squared error for the band interpolation along the path in
Figs.~\ref{fig:si.b34w30a}--\ref{fig:si.b34w30e} is (a) $0.192989$,
(b) $0.176692$, (c) $0.271211$, (d) $0.411963$, and (e) $0.695313$ eV$^2$.
Since the case of $\gamma=0.01$ is slightly better than the MLWF case of
$\gamma=0.0$ we use this value for the band structure shown in the main text
and previous section.

\subsection{Band structure interpolation and Brillouin zone sampling}

\begin{figure}[ht]
\subfloat[$6 \times 6 \times 6$ $\bk$-mesh; 12 bands (8 DLWFs)]{
    \includegraphics[width=0.45\textwidth]{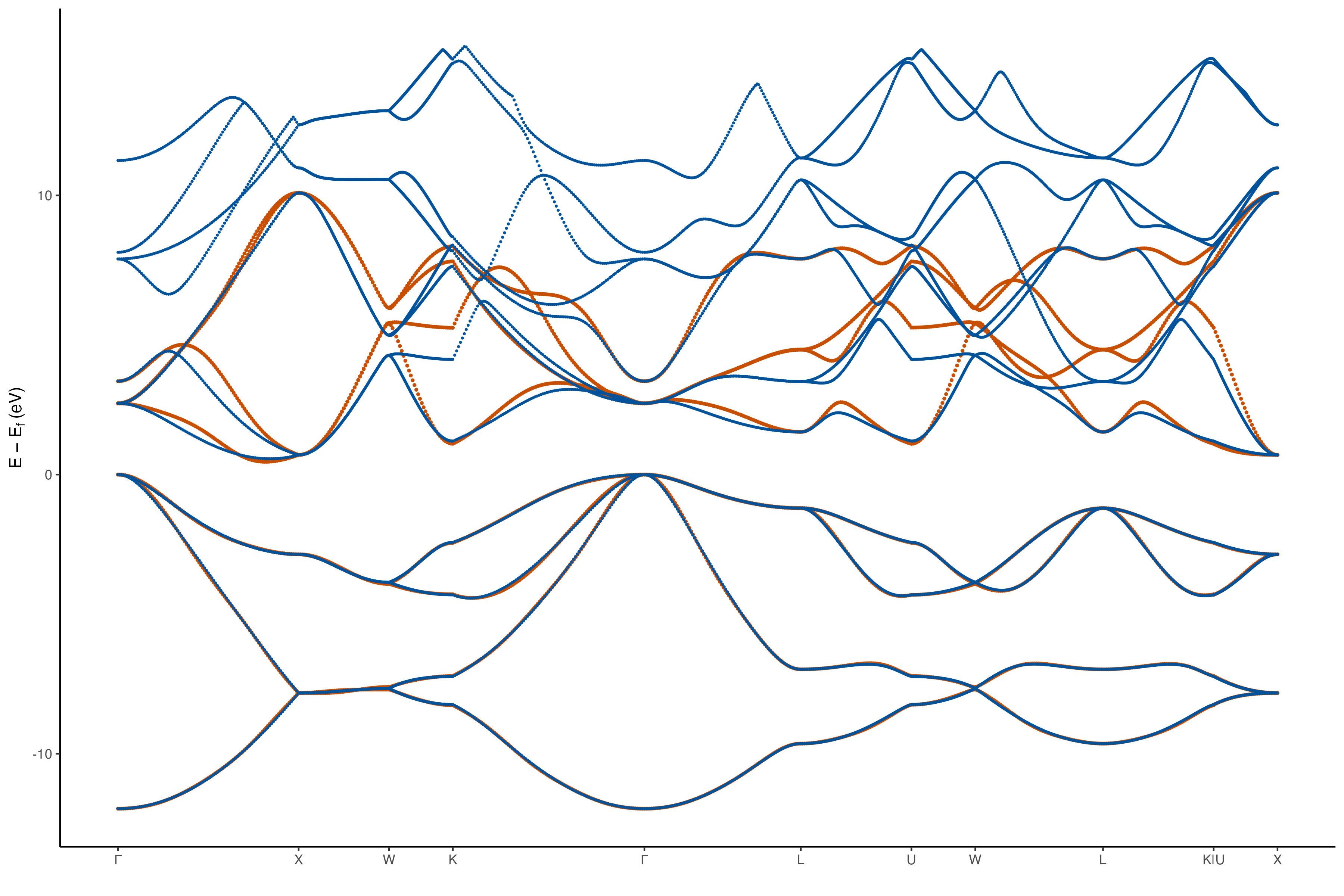}
    \label{fig:si-band_k06-b12}
}
\subfloat[$6 \times 6 \times 6$ $\bk$-mesh; 32 bands (28 DLWFs)]{
    \includegraphics[width=0.45\textwidth]{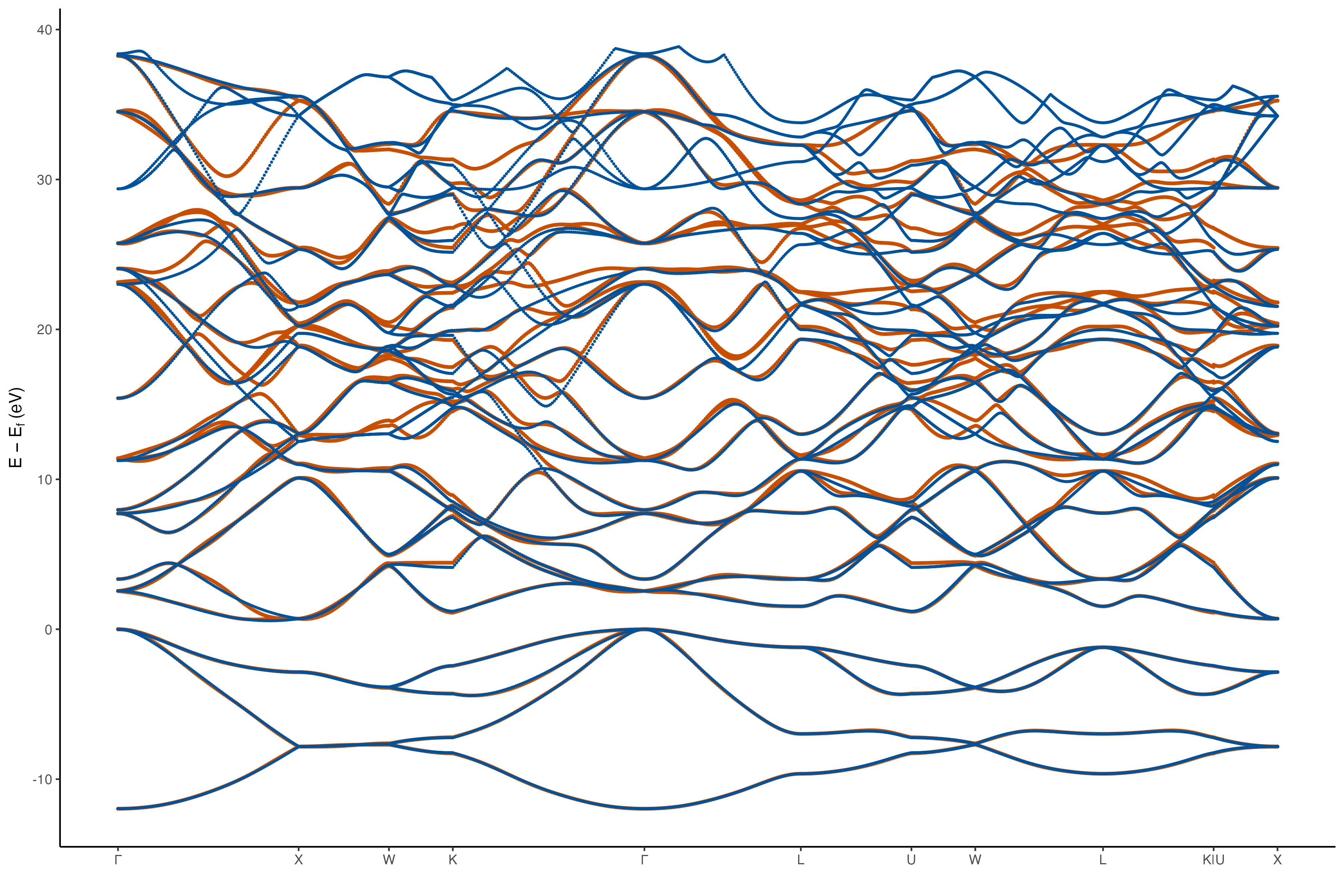}
    \label{fig:si-band_k06-b32}
} \\
\subfloat[$8 \times 8 \times 8$ $\bk$-mesh; 12 bands (8 DLWFs)]{
    \includegraphics[width=0.45\textwidth]{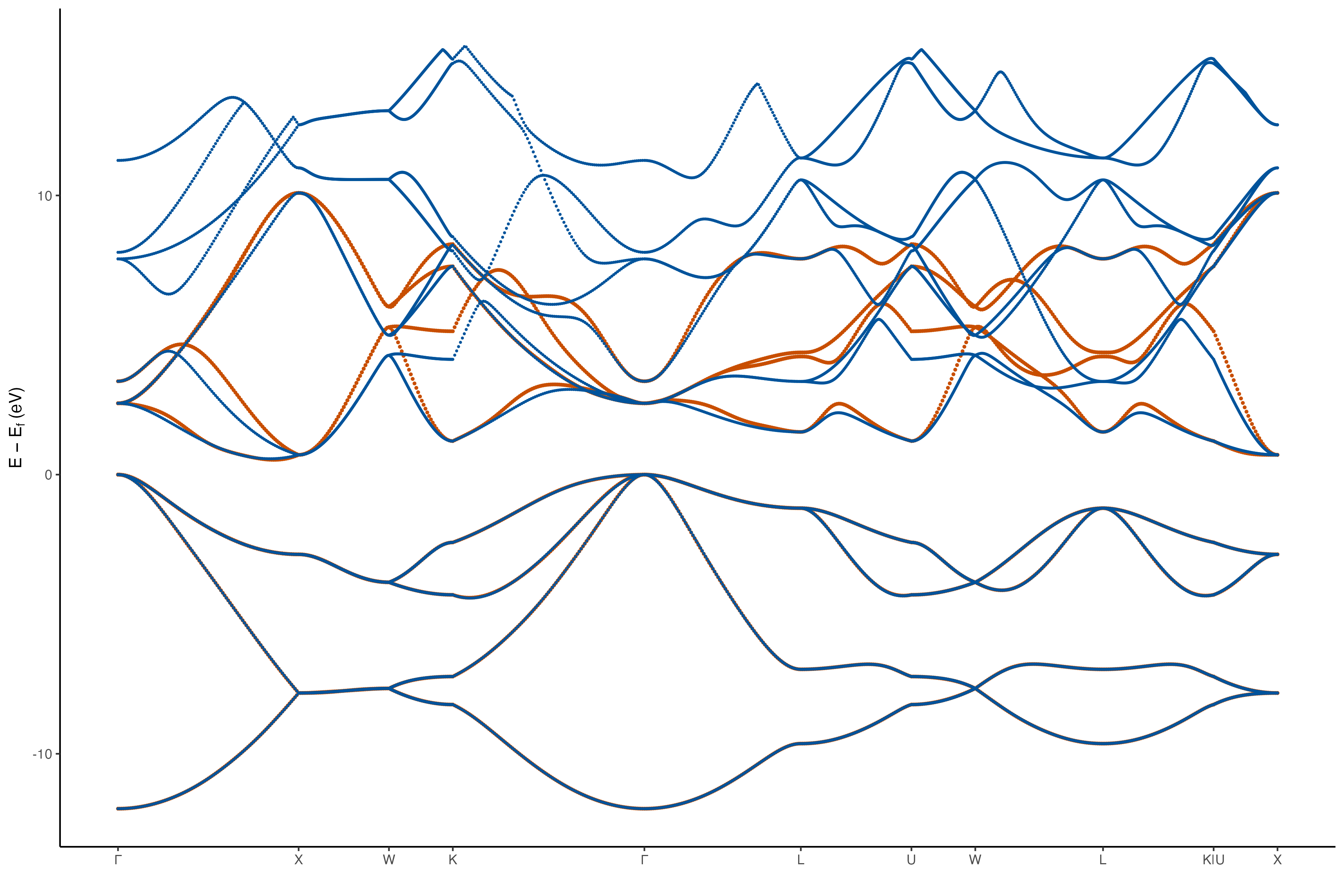}
    \label{fig:si-band_k08-b12}
}
\subfloat[$8 \times 8 \times 8$ $\bk$-mesh; 32 bands (28 DLWFs)]{
    \includegraphics[width=0.45\textwidth]{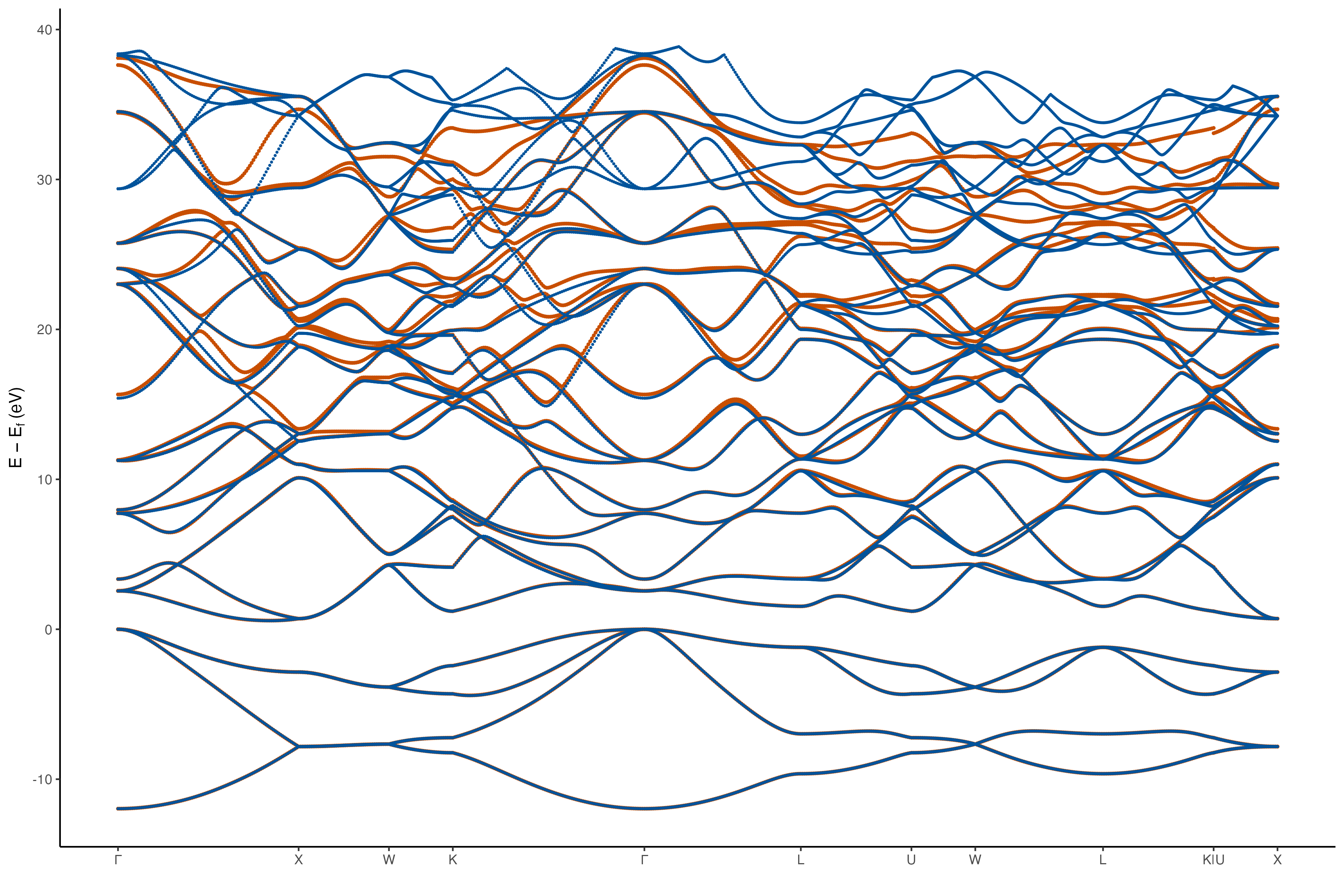}
    \label{fig:si-band_k08-b32}
} \\
\subfloat[$10 \times 10 \times 10$ $\bk$-mesh; 12 bands (8 DLWFs)]{
    \includegraphics[width=0.45\textwidth]{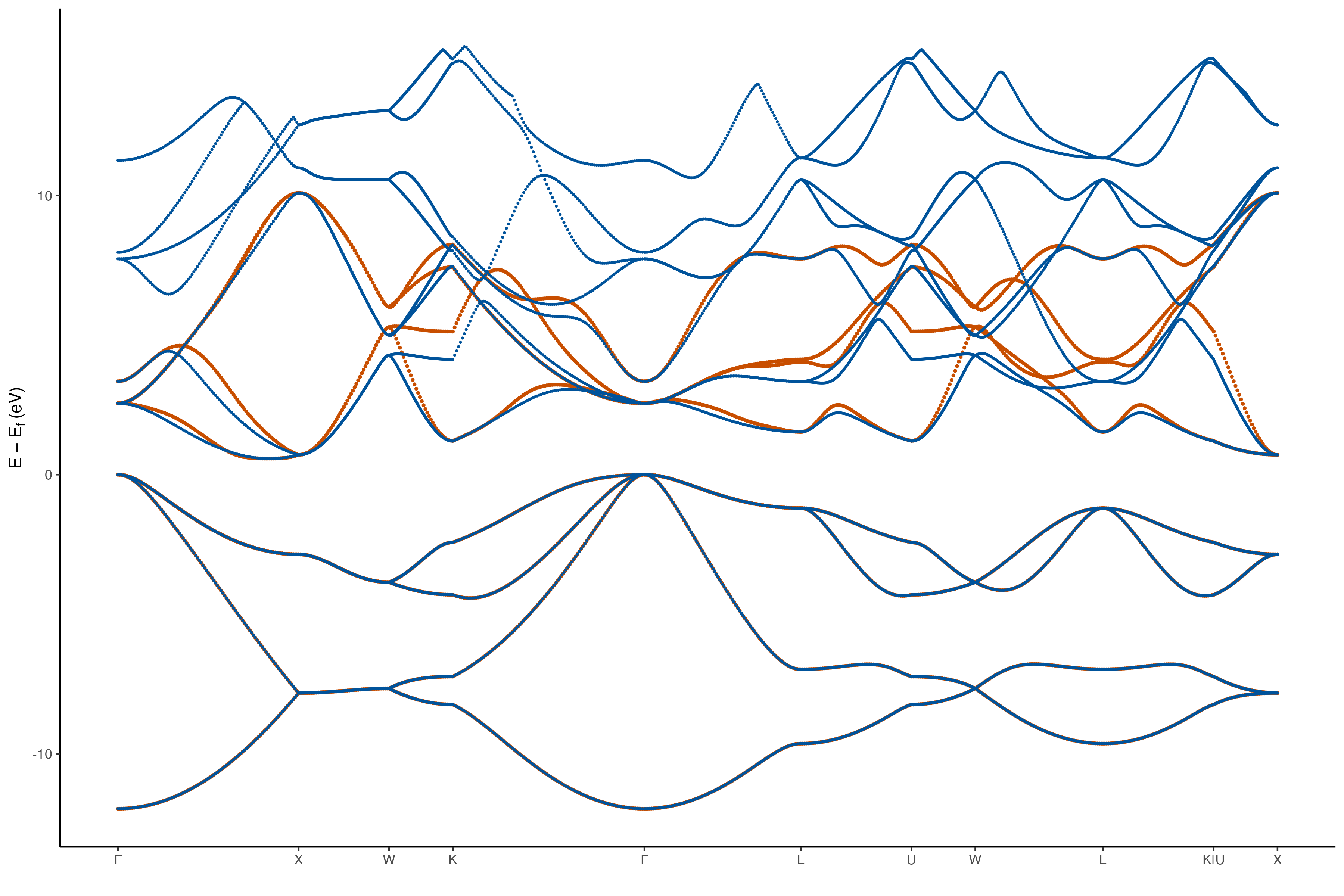}
    \label{fig:si-band_k10-b12}
}
\subfloat[$10 \times 10 \times 10$ $\bk$-mesh; 32 bands (28 DLWFs)]{
    \includegraphics[width=0.45\textwidth]{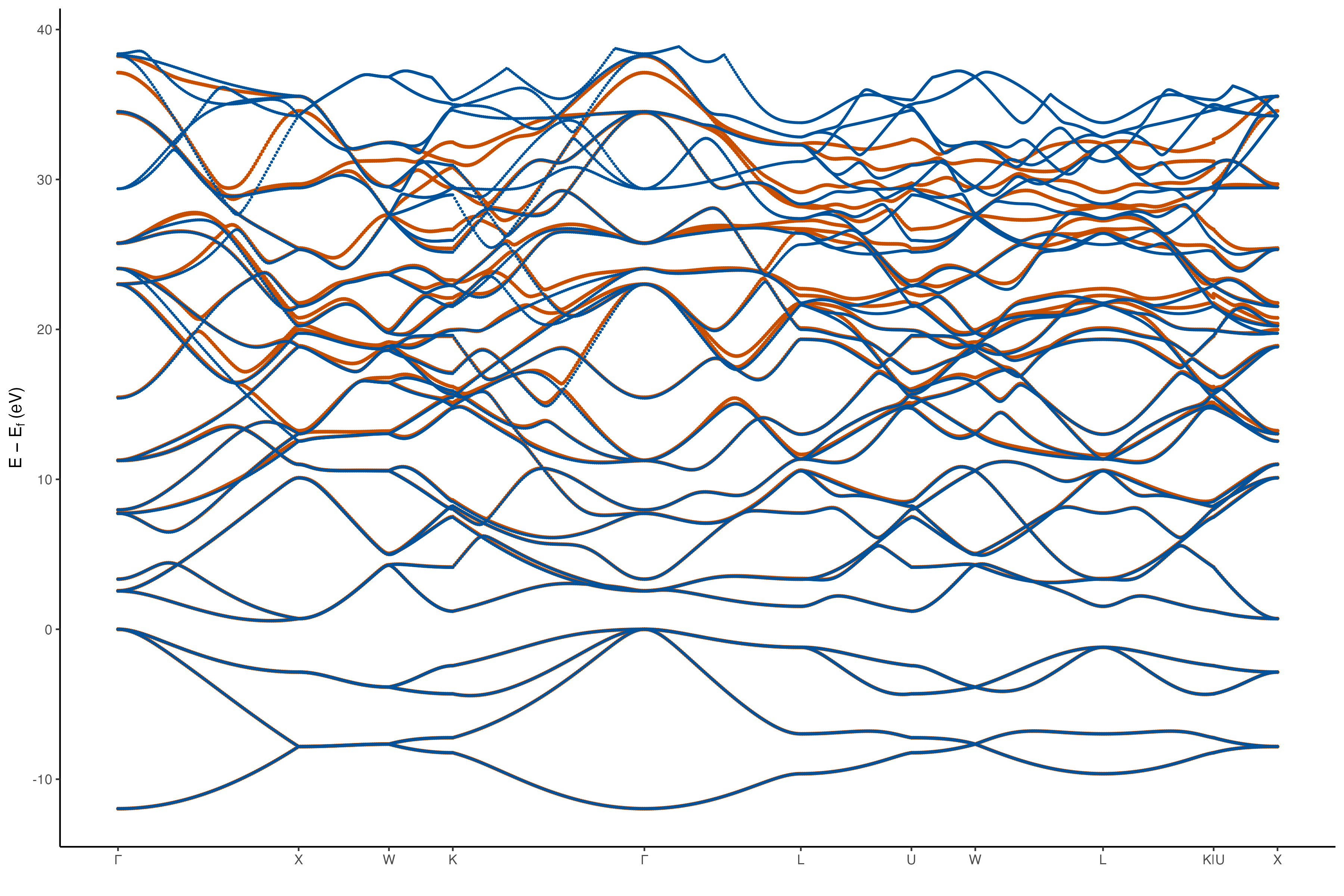}
    \label{fig:si-band_k10-b32}
}
\caption{Band structure of silicon computed from the DFA (blue) and 
         interpolated from the DLWFs (orange). The valence band maximum is set
         to zero. Subfigures have different Monkhorst--Pack samplings of the
         Brillouin zone and different numbers $N_b$ of bands from which the
         $N_w$ DLWFs are constructed ($N_w = N_b - 4$).}
\label{fig:si-bands-k}
\end{figure}

With 12 bands disentangled to 8 DLWFs (the left column of
Fig.~\ref{fig:si-bands-k}), we do not reconstruct the conduction bands
quantitatively because they are affected directly by the disentanglement
procedure. The same is true of the highest-energy conduction bands in the
right column, but they are far above the Fermi energy. Regardless, the frontier
bands are reproduced fairly accurately. Some oscillations in the low-lying DLWF
conduction bands, relative to the DFA bands, can be seen even with 28 DLWFs
(Fig.~\ref{fig:si-band_k06-b32}); this is ameliorated for larger
$\bk$-samplings. These oscillations do affect the DLWF-interpolated prediction
for the conduction band minimum in silicon, which does not lie on a
high-symmetry $\bk$-point. When it is critical to reproduce the (DFA) conduction
band minimum accurately, a denser sampling is thus recommended.

\clearpage

\section{Mixing Parameter Conversion}
In the molecular formulation of the localized orbital scaling correction (LOSC)
method \cite{su2020}, the cost function is given as
\begin{equation} \label{eqn:l2cost_mol}
    F = (1-\gamma')\Omega + \gamma' C' \Xi,
\end{equation}
where $\Omega$ is the spatial variance cost and $\Xi$ is the energy variance
cost. We write $\gamma'$ to differentiate from the mixing parameter $\gamma$
used in the main text. The constant $C'$ is added because in atomic units there
is a large order of magnitude difference between typical space and energy
spreads. This results in mixing parameters that are very close to $1.0$, which
are difficult to optimize. To keep the space and energy cost within the same order
of magnitude, the factor of $C$ was set to $\qty{1000}{\bohr^2/\hartree^2}$
(where \unit{\bohr} is the Bohr radius and \unit{\hartree} is one hartree).
The implementation in {\tt wannier90} uses
the units of \unit{\angstrom} and \unit{\electronvolt},
which results in spreads that are of the same order of magnitude; thus,
we use a factor of $C = \qty{1}{\angstrom^2/\electronvolt^2}$ in our 
implementation of the cost function. Converting between these two
different equations and units can be accomplished with the following formulae.
When converting from Eq.~\eqref{eqn:l2cost_mol}, the equivalent mixing parameter
for the cost function in the main text is given by
\begin{equation}
    \gamma = \Bigg( 1 + \left(\frac{1-\gamma'}{\gamma'}\right) m^2 C' \Bigg)^{-1}
\end{equation}
where $m = 0.529177210903/27.211386245988$ is the conversion from
\unit{\bohr} to \unit{\angstrom} divided by that from \unit{\hartree} to
\unit{\electronvolt}. Values are obtained from the NIST
CODATA database \cite{tiesinga2020}. The conversion from the mixing parameter in
the main text to the cost function in Eq.~\eqref{eqn:l2cost_mol} is given by 
\begin{equation}
    \gamma' = \Bigg( 1 + \left(\frac{1-\gamma}{\gamma m^2 C'}\right) \Bigg)^{-1}.
\end{equation}

\section{Code Implementation Details}
The cost function outlined in the main text was implemented in the
{\tt wannier90} code \cite{mostofi2008, mostofi2014, pizzi2020}. To implement
the energy localization, the data structures necessary for energy localization
are only calculated if energy localization is requested. The energy spread cost
is calculated in the function {\tt wann\_xi}, which is called right after the
function {\tt wann\_omega}, which calculates the spatial spread. Similarly, the
energy gradient is only calculated when needed and is calculated in
{\tt wann\_dxi}, which is called right after the function that calculates the
spatial spread gradient, {\tt wann\_domega}.  The analytic gradient implemented
in {\tt wann\_xi} was verified numerically for some test systems to ensure the
derivation is correct. The additional data structures that hold the pertinent
information for energy localization are {\tt have} for average energy,
{\tt have2} for average energy squared, and {\tt h2ave} for average of the
squared energy. These variables are used to update the public variables
{\tt wannier\_energies} and {\tt wannier\_espreads}, which are the energy version
of the public variables {\tt wannier\_centres} and {\tt wannier\_spreads} used for
spatial information. Since adding these variables into the library mode would
necessitate a change to the API, this change was not implemented. If desired,
these variables could be added to the API as optional arguments, which should
prevent breaking any existing code that uses the API. Our modified code is hosted
on GitHub at \cite{mahler2024a}.

\subsection{Keyword List}
The {\tt wannier90} program uses an input file named {\tt seedname.win}, where
{\tt seedname} can be set from the command line. This input file contains
key-value pairs in a free-form structure that controls the settings of the
program. This is a list of the new keywords added in the implementation of
energy localization in the {\tt wannier90} code.

\begin{longtable}{|c|c|p{6cm}|}
  \hline
  Keyword & Type & Description \\
          &      &             \\
  \hline\hline
  \multicolumn{3}{|c|}{Energy Mixing Parameters \vphantom{{$\big()$}}} \\
  \hline
  {\sc sp\_en\_mix }                & R & Space and energy mixing parameter \\
  {\sc econv\_max }                 & R & Energy convergence maximum \\
  {\sc nconv\_max }                 & I & Number of WFs to use for convergence \\
  {\sc num\_occ }                   & I & Bloch orbital occupation per k-point \\
  {\sc write\_info }                & L & Write localization information per WF to file \\
  {\sc wannier\_plot\_density }     & L & Plot the WF density \\
\hline
\addlinespace[2ex]
\caption[Parameter file keywords controlling energy localization.]
{{\tt seedname.win} file keywords controlling the  energy localization.  Argument types
are represented by, I for a integer, R for a real number, and L for a logical value.}
\label{parameter_keywords6}
\end{longtable}

\subsection{Keyword Details}

\subsubsection*[sp\_en\_mix]{\tt real :: sp\_en\_mix}
Space and energy mixing parameter; a value of $0.0$ gives the MLWF cost function.
This value must obey the inequality $ 0.0 \leq \text{{\tt sp\_en\_mix}} \leq 1.0$.
When $ 0.0 < \text{{\tt sp\_en\_mix}}$ then the per WF information printed at
every print cycle will change to include the energy statistics.

The default value is $0.0$.

\subsubsection*[econv\_max]{\tt real :: econv\_max}
The upper bound for the energy window used for testing localization convergence.
This is useful when considering high-energy unoccupied orbitals since they can be
very noisy during the descent.  If {\tt econv\_max} is set then an additional line
will be printed at the end of each print cycle showing the spatial, energy, and
total cost of the subset used for convergence.  This setting cannot be set if
{\tt nconv\_max} is set.

No default.

\subsubsection*[nconv\_max]{\tt integer :: nconv\_max}
The number of Wannier functions used to test for convergence during the gradient
descent in the Wannierisation procedure. Useful for when considering high-energy
unoccupied orbitals since they can be too noisy to converge during the descent.
If {\tt nconv\_max} is set then an additional line will be printed at the end of
each print cycle showing the spatial, energy, and total cost of the subset used
for convergence.  This setting cannot be set if {\tt econv\_max} is set.

No default, but if {\tt nconv\_max} $>$ {\tt num\_wann} it will be set to
{\tt num\_wann}.

\subsubsection*[num\_occ]{\tt integer :: num\_occ}
The number of occupied of Bloch orbitals. This is used to print the occupation
of the WFs. Since it is only a single integer this implementation only applies
correctly for insulators and $\Gamma$-point calculations. In the typical case of
finding MLWFs for occupied only orbitals the WF occupations will always be
$1.0$. Setting {\tt num\_occ} is useful when considering WFs that include both
occupied and unoccupied Bloch orbitals.

No default.

\subsubsection*[write\_info]{\tt logical :: write\_info}
If {\tt true}, write a summary of the cost information per WF to the file
{\tt seedname.info} for the WFs in the home unit cell. The format is WF index,
Cartesian expectation of center in \AA, spatial spread in \AA$^2$. If
{\tt sp\_en\_mix} $\neq 0.0$ then it will additionally print the WF energy
expectation in eV and the energy spread in eV$^2$. If {\tt num\_occ}
$\neq 0$ then it will also print the WF occupations. 

The default value is {\tt false}.

\subsubsection*[wannier\_plot\_density]{\tt logical :: wannier\_plot\_density}
Plot the WF density instead of the wavefunction, where the density is defined as
$\vert w_n(\mathbf{r})\vert^2$. In the case of WFs that are not strictly
real-valued, plotting the wavefunction will truncate the imaginary part.
Plotting the density instead guarantees the whole WF is plotted.

The default value is {\tt false}.

\bibliography{DLWF}